\documentclass[apj, iop]{emulateapj}
\usepackage{gensymb}

\usepackage{natbib,twoopt}
\makeatletter
\newcommandtwoopt{\citeads}[3][][]{\href{http://adsabs.harvard.edu/abs/#3}%
{\def\hyper@linkstart##1##2{}%
\let\hyper@linkend\@empty\citealp[#1][#2]{#3}}}
\newcommandtwoopt{\citepads}[3][][]{\href{http://adsabs.harvard.edu/abs/#3}%
{\def\hyper@linkstart##1##2{}%
\let\hyper@linkend\@empty\citep[#1][#2]{#3}}}
\newcommandtwoopt{\citetads}[3][][]{\href{http://adsabs.harvard.edu/abs/#3}%
{\def\hyper@linkstart##1##2{}%
\let\hyper@linkend\@empty\citet[#1][#2]{#3}}}
\newcommandtwoopt{\citeyearads}[3][][]%
{\href{http://adsabs.harvard.edu/abs/#3}
{\def\hyper@linkstart##1##2{}%
\let\hyper@linkend\@empty\citeyear[#1][#2]{#3}}}
\makeatother

\begin{document}

\title{Properties of supersonic Evershed downflows}
\author{S. Esteban Pozuelo$^{1,2}$, L.~R. Bellot Rubio$^{1}$ \and J. de la Cruz Rodr\'{i}guez$^{2}$}

\affil{$^{1}$Instituto de Astrof\'{i}sica de Andaluc\'{i}a (CSIC), Glorieta de la Astronom\'{i}a s/n, 18080 Granada, Spain}
\affil{$^{2}$Institute for Solar Physics, Dept. of Astronomy, Stockholm University, AlbaNova University Center, 106 91 Stockholm, Sweden, \\ current email: sara.esteban@astro.su.se}

  \begin{abstract}
We study supersonic Evershed downflows in a sunspot penumbra by means of high spatial resolution spectropolarimetric data acquired in the \ion{Fe}{1}~617.3~nm line with the CRISP instrument at the Swedish 1-m Solar Telescope. Physical observables, such as Dopplergrams calculated from line bisectors and Stokes~$V$ zero-crossing wavelengths, and Stokes~$V$ maps in the far red wing, are used to find regions where supersonic Evershed downflows may exist. We retrieve the LOS velocity and the magnetic field vector in these regions using two-component inversions of the observed Stokes profiles with the help of the SIR code. We follow these regions during their lifetime to study their temporal behavior. Finally, we carry out a statistical analysis of the detected supersonic downflows to characterize their physical properties. Supersonic downflows are contained in compact patches moving outward, which are located in the mid and outer penumbra. They are observed as bright, roundish structures at the outer end of penumbral filaments that resemble penumbral grains. The patches may undergo fragmentations and mergings during their lifetime, even some of them are recurrents. Supersonic downflows are associated with strong and rather vertical magnetic fields with a reversed polarity compared to that of the sunspot. Our results suggest that downflows returning back to the solar surface with supersonic velocities are abruptly stopped in dense deep layers and produce a shock. Consequently, this shock enhances the temperature and is detected as a bright grain in the continuum filtergrams, which could explain the existence of outward moving grains in the mid and outer penumbra.
\end{abstract}


\section{Introduction}
\label{sec:introduction}

The penumbra harbors a plethora of mass motions due to the presence of magnetoconvection in sunspots \citepads{1965ApJ...141..548D}. Among the most intriguing and less understood are supersonic\footnote{The sound speed in the photosphere is 7.2~km~s$^{-1}$.} downflows.

The existence of strong Evershed downflows in the penumbra is suspected since long. They were first reported by \citet{bumba60} as line flags, where the line is not shifted as a whole but instead has a very strong asymmetry (flag). Such extreme cases of line asymmetries were interpreted in terms of a superposition of two unresolved structures: an almost unshifted strong component, which corresponds to the "mean" Evershed effect \citepads{1909MNRAS..69..454E}, and a weaker strongly displaced component (satellite) where the line-of-sight (hereafter, LOS) velocity reaches up to 7-8~km~s$^{-1}$ (\citeads{1960IAUS...12..403S}, \citeads{1961AnAp...24....1S};~\citeads{1964ApNr....8..205M};~\citeads{1971SoPh...18..220S};
~\citeads{1995A&A...298L..17W}). 

In the early 2000s, inversion codes allowed a much more precise determination of flows in sunspot penumbrae through the use of full Stokes line profiles. \citetads{2001ApJ...549L.139D} and ~\citetads{2004A&A...427..319B}, for example, inferred the existence of supersonic downflows from the inversion of spectropolarimetric measurements in the infrared showing Stokes~$V$ profiles with strong asymmetries and multiple lobes. However, the downflows could not be imaged directly due to insufficient spatial resolution of the observations (about 1\arcsec). The advent of Hinode satellite \citepads{2007SoPh..243....3K}  brought spectropolarimetric data with improved spatial resolution (0\farcs32). Using a set of magnetograms in the red far wing of \ion{Fe}{1}~630.2~nm, \citetads{2007PASJ...59S.593I} found the sinks of the Evershed flow and circular polarization measurements enter the scene. According to these authors, the bisector technique yields LOS velocities of 4-7~km~s$^{-1}$ in such sinks. Thereafter, \citetads{2010ASSP...19..193B} was the first to show a reversed two-lobed Stokes~$V$ profile associated to a LOS velocity of about 9 km~s$^{-1}$ (in concordance with those obtained by \citetads{2009A&A...508.1453F} from bisectors in pixels with rare circular polarization signals). Independently to the finding of \citetads{2010ASSP...19..193B}, the spatial resolution acquired by the Hinode satellite is not high enough to spatially resolve structures harboring supersonic downflows and Stokes~$V$ profiles are usually irregular, showing 3 or 4 lobes. In this regard, \citetads{2013A&A...557A..24V} has been the first to infer the LOS velocity and the magnetic field vector associated to supersonic downflows from spatially-coupled inversions \citepads{2012A&A...548A...5V}. However, they inferred extreme LOS velocity and magnetic field strength values in some cases (around 20~km~s$^{-1}$ and 7~kG, respectively). Furthermore, the Stokes profiles represented in their Figure~3 show very weak linear polarization signals and a very complex (pathological) Stokes~$V$ profile that has been inverted using only a component. 

\begin{figure*}[!t]
\centering
\includegraphics[trim = {3cm 0.5cm 3cm 1cm}, clip, width = 0.85\textwidth]{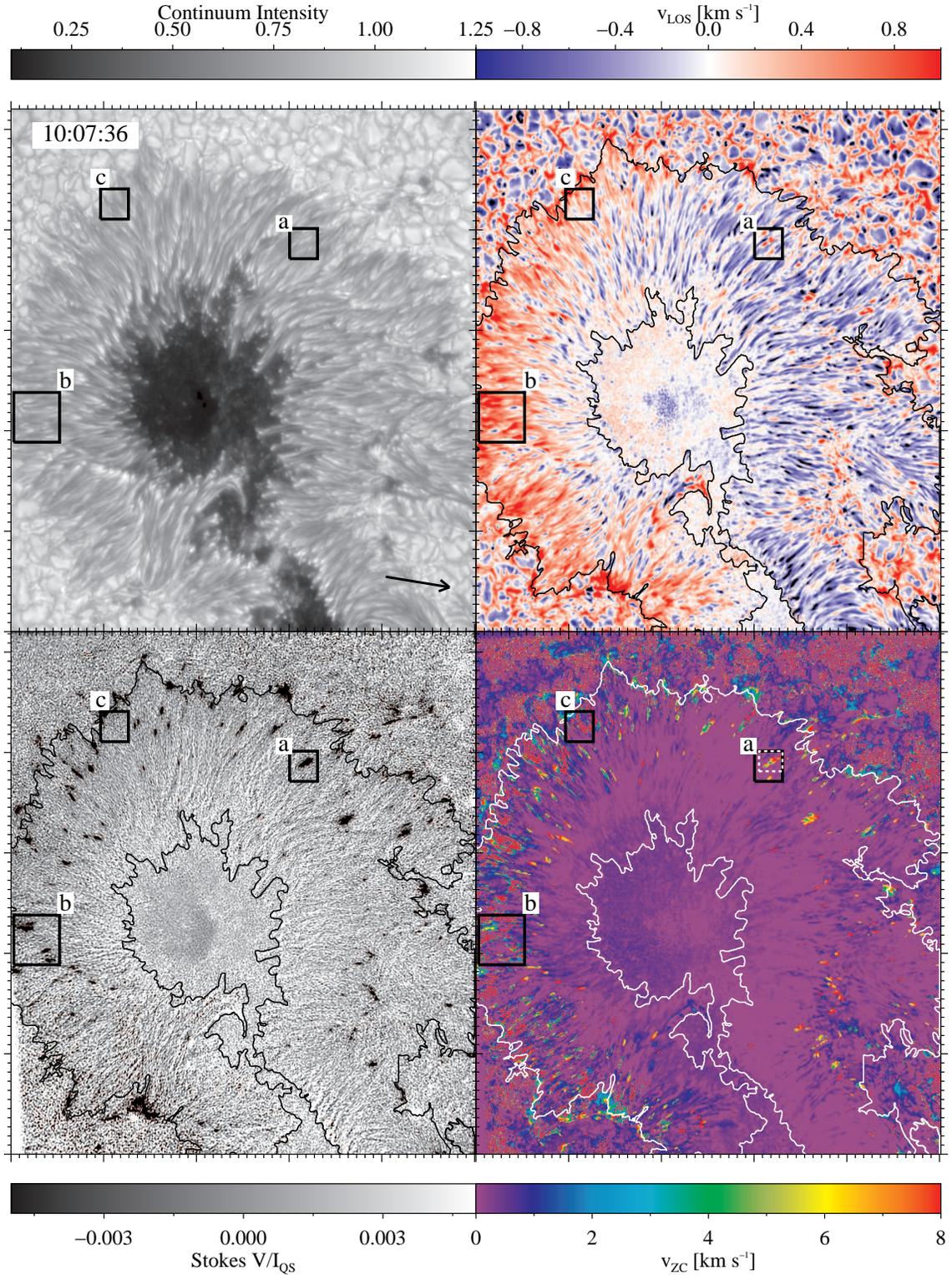} 
\caption{Top row: continuum intensity image of the main sunspot of AR 11302 observed on 28 September 2011, 10:07:36~UT (left panel) and Dopplergram as derived from the line bisectors at the 70\% intensity level (right panel). Bottom row: the corresponding Stokes~$V$ map at +420~m\AA\, and the LOS velocity map given by the Stokes~$V$ zero-crossing points. Squares enclose three regions that will be studied in Section~\ref{sub:temporal}. The contours outline the penumbral region. The black arrow points to the disk center. Each major tickmark represents 10\arcsec.}
\label{fig:intensity_observables}
\end{figure*}

Despite all previous efforts, we still do not have direct proofs of the existence of supersonic downflows in the penumbra. We do not know if they leave spectral signatures on all four Stokes profiles, or only in Stokes~$I$ and~$V$. Moreover, an important (and missing) element is the temporal evolution: to know how they appear, evolve and disappear is essential to understand the nature of supersonic downflows.

Our motivation is entirely focused on the aspects mentioned above. Specifically, we study in-depth for the first time the morphology and temporal evolution of supersonic downflows in a sunspot penumbra, as well as characterize their physical properties. This goal has been achieved thanks to a time sequence of high spatial resolution (0\farcs13) and high cadence spectropolarimetric data of the \ion{Fe}{1}~617.3~nm spectral line acquired with the CRISP instrument at the Swedish 1-m Solar Telescope under excellent seeing conditions.

This paper is organized as follows. The observations and the data reduction process are briefly described in Section~\ref{sec:observations_datareduction}. This is followed by Section~\ref{sec:dataanalysis} where we define the proxies used to detect areas susceptible of supersonic Evershed downflows. In Section~\ref{sec:supersonic}, we characterize the shape and the spatial distribution of the Stokes profiles emerging from patches harboring supersonic downflows and, finally, we describe how inversions were performed. In Section~\ref{sec:results}, we outline the temporal behavior of supersonic Evershed downflows by analyzing the temporal evolution of three spatially-resolved examples. After that, the physical properties of the detected supersonic downflows are enumerated. These results are discussed and compared with previous studies (Section~\ref{sec:discussion}). Finally, Section~\ref{sec:conclusion} offers a summary of our results and a conclusion.

\section{Observations and data reduction}
\label{sec:observations_datareduction}

We observed the main sunspot of active region 11302 under excellent seeing conditions on September 28, 2011 between 09:20:40 and 10:10:40~UT (upper left panel of Figure~\ref{fig:intensity_observables}). The spot was located very close to the disk center, at an heliocentric angle of~6.8$\degree$. Therefore, this sunspot is suitable for studying vertical gas motions in penumbrae, since projections effects are minimized. 

We recorded a time series of full-Stokes measurements in the \ion{Fe}{1}~617.3~nm line with the CRISP spectropolarimeter (\citeads{2006A&A...447.1111S};~\citeads{2008ApJ...689L..69S}) at the Swedish 1-m Solar Telescope \citepads[SST,][]{2003SPIE.4853..341S}. The spectral sampling consists of 30 wavelength positions, from $-$35.0~to~$+$66.5~pm in steps of 3.5~pm. This scanning allows us to detect strong redshifts present in the clean continuum of the spectral line. Each scan required 32~s to be completed. The duration of this serie is $\sim$49~minutes (98~frames).

The data were reduced using the CRISPRED pipeline \citepads{2015A&A...573A..40D} and processed with the Multi-Object-Multi-Frame-Blind-Deconvolution technique (MOMFBD,~\citeads{1994A&AS..107..243L};~\citeads{2005SoPh..228..191V}). The polarimetric calibration was performed for each pixel of the FOV as suggested by \citetads{2008A&A...489..429V}. The same dataset was used in a previous paper \citepads{2015ApJ...803...93E} where more details about the observations and the data reduction process can be found.

\section{Data analysis}
\label{sec:dataanalysis}

In this section, we describe the parameters used to detect signatures of supersonic Evershed flows in the observed Stokes profiles. According to previous studies (Section~\ref{sec:introduction}), asymmetries and shifts in Stokes~$I$ and~$V$ profiles suggest the presence of supersonic Evershed downflows. Therefore, we have used Dopplergrams obtained using line bisectors and Stokes~$V$ zero-crossing points, and magnetograms taken in the far red wing as proxies to detect them.

\begin{figure}[!t]
\centering
\includegraphics[trim = {2cm 2cm 2cm 2cm}, clip, width = 0.50\textwidth]{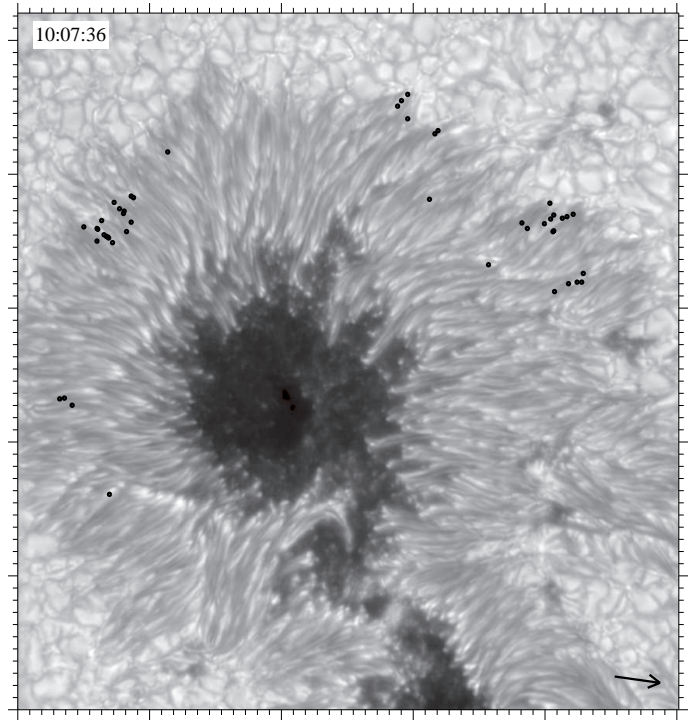} 
\caption{Location of the 40 supersonic patches overplotted on the intensity map shown in Figure~\ref{fig:intensity_observables}. The black arrow points to the disk center. Each major tickmark represents 10\arcsec.}
\label{fig:distribucion}
\end{figure}

\begin{figure*}[!t]
\centering
\includegraphics[trim = {1.5cm 1cm 3.25cm 3cm}, clip, width = 0.55\textwidth]{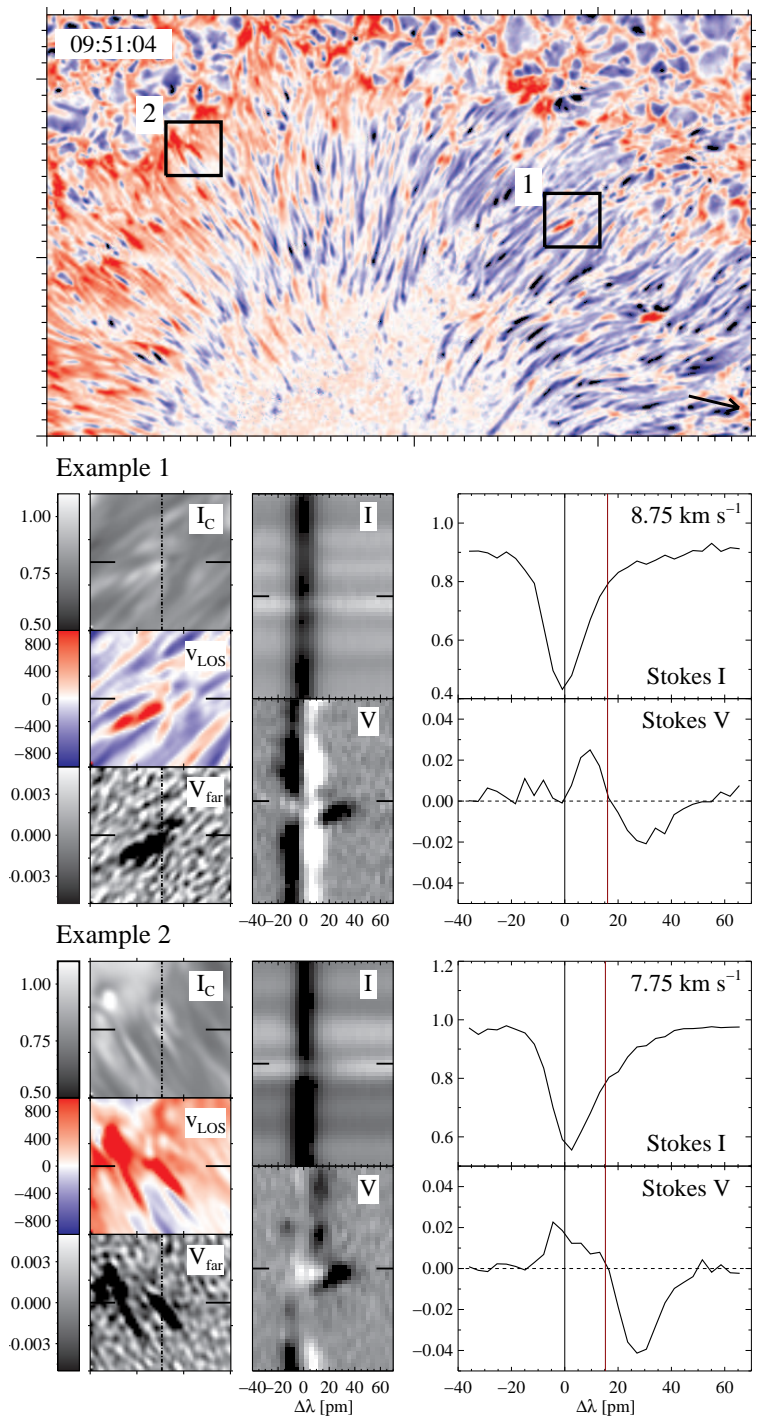} 
\caption{Examples of supersonic Evershed downflows detected in AR 11302. The upper half of a Dopplergram acquired at 09:51:04~UT shows two cases enclosed by squares (top panel). Each major tickmark represents 10\arcsec. The black arrow indicates the direction to the disk center. These two examples are described in the central and bottom panels, respectively. For each case we display close-ups of the continuum intensity, Dopplergram and magnetogram in the far red wing (leftmost panels, each tickmark represents 1\arcsec), Stokes~$I$ and~$V$ spectra emerging along the dashed line drawn in the intensity map and magnetogram (center panels), and Stokes~$I$ and~$V$ profiles observed at the position of the supersonic downflow marked with horizontal dashes on both sides of each panel.}
\label{fig:ejemplos_generales}
\end{figure*}

The upper right panel of Figure~\ref{fig:intensity_observables} shows a LOS velocity map derived from line bisectors as in \citetads{2015ApJ...803...93E}, where they were calculated from observed intensity profiles using linear interpolation (after the spectral gradients caused by the CRISP prefilter were removed). The Evershed flow is stronger close to the continuum forming layer (e.g.,~\citeads{1995A&A...298..260R};~\citeads{2003A&A...403L..47B}), so we computed Dopplergrams from the line bisectors at the 70\% intensity level\footnote{The core and the continuum define the 0 and 100\% intensity levels, respectively.}. Furthermore, following \citetads{1989ApJ...336..475T}, we applied a Fourier filter with a cut off speed of 5~km~s$^{-1}$ to remove the subsonic oscillations \citepads{1962ApJ...135..474L} from our LOS velocity maps. The LOS velocity was calibrated using the zero-crossing wavelength of symmetric Stokes~$V$ profiles in the umbra to avoid possible molecular blends in the darkest umbral positions \citepads{2006SoPh..239...69N}. The typical standard deviation of the velocity reference is 110~m~s$^{-1}$.

In the Dopplergram we see that strong redshifts concentrate in areas located at the end of flow channels, which indicate positions where the Evershed flow is returning back to the solar surface. These redshifts are usually found in the mid and outer penumbra, and weaker in inner regions. Although they are most easily detected in the center side penumbra because their contrast is higher there, they also appear in the limb side penumbra. According to \citetads{2010ASSP...19..193B} and \citetads{2013A&A...557A..24V}, these flows are supersonic and nearly vertical. At those position, the bisector technique yields moderate LOS velocities of about 2~km~s$^{-1}$. Dopplergrams are computed considering only Stokes~$I$ profiles, but supersonic Evershed downflows do not show up clearly in the intensity profiles. Therefore, we have to examine the Stokes~$V$ profiles to see them better.

\begin{figure}[!t]
\centering
\includegraphics[trim = {0.25cm 0.75cm 2cm 2.75cm}, clip, height = 0.85\textheight]{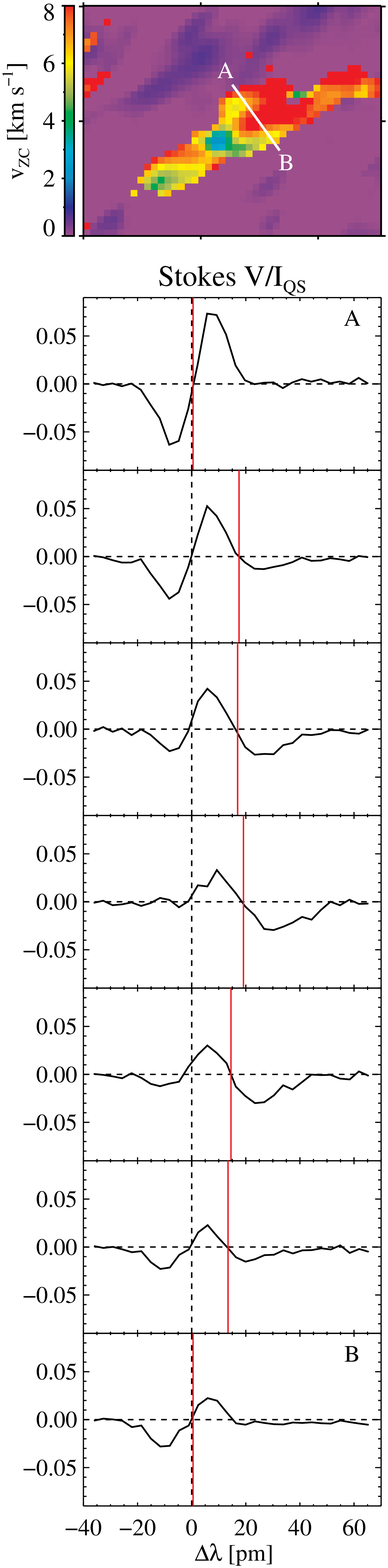} 
\caption{Stokes V profiles observed in the mid center-side penumbral region labeled a in Figure~\ref{fig:intensity_observables}. The upper panel shows a close-up of the velocity map. Each tickmark indicates 1\arcsec. The Stokes~$V$ profiles of different pixels along the cut drawn in the close-up are represented in the other panels. The x-axis contains the relative wavelength position from 617.33~nm. The red vertical lines indicate the Stokes~$V$ zero-crossing wavelengths.}
\label{fig:parche_stokesv}
\end{figure}

Following \citetads{2007PASJ...59S.593I}, we have inspected the Stokes~$V$ filtergrams taken in the far red wing of \ion{Fe}{1}~617.3 nm at $+$420~m\AA\footnote{This wavelength position corresponds to a LOS velocity of about 22~km~s$^{-1}$.}\, to search for positions with enhanced signal that might contain supersonic Evershed downflows. An example is shown in the lower left panel of Figure~\ref{fig:intensity_observables}. Prominent patches are observed in the mid and outer penumbra. Their presence may be due to strong redshifts (Doppler effect), strong magnetic fields (Zeeman splitting), or a combination of both. Therefore, these magnetograms can not be used as an unique proxy to detect supersonic downflows, since not all patches are actually due to strong redshifts.

To narrow down the search we have checked whether the Stokes~$V$ profiles are strongly redshifted. Thus, we have also computed LOS velocity maps from the Stokes~$V$ zero-crossing points, and applied the same velocity calibration as in the Dopplergrams. We show an example in the lower right panel in Figure~\ref{fig:intensity_observables}, where only redshifted LOS velocities are represented. Thus, according to the color table, supersonic downflows might occur in the orange-red elliptical patches that are located in different penumbral regions. 

We have combined all these proxies to get an idea of the Stokes~$I$ and~$V$ shifts and relate them to the existence of supersonic Evershed flows. This has allowed us to find 40~patches containing supersonic Evershed flows whose initial positions are marked with black circles in Figure~\ref{fig:distribucion}. These areas are usually located in the outer penumbral regions on both sides of the sunspot, but some of them are also found in the mid penumbra. Despite we have also examined inner penumbral positions, we do not find any supersonic Evershed flow there. We focus on ten of them that show a regular behavior over time and do not have very irregular Stokes profiles. They represent around 0.16\%\, of the penumbral pixels.

\section{Supersonic Evershed flows}
\label{sec:supersonic}

In this section we examine the profiles generated by supersonic Evershed flows and study how they are distributed within the patches.

The top panel of Figure~\ref{fig:ejemplos_generales} shows two regions where the bisector velocity at the 70\%\, level is strongly redshifted. Some aspects of these examples are displayed in detail in the lower part of Figure~\ref{fig:ejemplos_generales}. 

Example 1 in Figure~\ref{fig:ejemplos_generales} corresponds to a strong downflow observed in the mid region of the center side penumbra\footnote{The temporal sequence of this example is described in Section~\ref{sub:examplea}.}. In the left column, from top to bottom, we show a continuum intensity filtergram, a Dopplergram and a far red-wing magnetogram over a 3\arcsec\,x 3\arcsec\, area containing the downflow. The two horizontal dashes on both sides of each panel mark the position of the downflow. The strong downflow coincides with a bright intensity feature and enhanced signals in the magnetogram. In the middle column, we display the Stokes~$I$ and~$V$ spectra (top and bottom panel, respectively) observed along the vertical dashed line drawn in the intensity map and in the magnetogram. At the position of the downflow, the continuum of the Stokes~$I$ profile is brighter than elsewhere and the Stokes~$V$ profile has reversed polarity and is strongly redshifted as a whole. The supersonic profiles are plotted in the right panels. The red wing of the Stokes~$I$ profile is very asymmetric and extended, but the profile is relatively unshifted, which results in small bisector velocities. By contrast, the Stokes~$V$ profile has two lobes that appear reversed in sign and are strongly redshifted. The Stokes~$V$ zero-crossing point is located at around $+$16~pm relative to the rest position. Although we cannot relate this shift to a homogeneous Doppler velocity (since the Stokes~$V$ profile has asymmetries), it corresponds to a velocity of 8.75~km~s$^{-1}$. Example 2 is located at the outer boundary of the limb side penumbra and shows similar characteristics as the previous one\footnote{Its temporal evolution is shown in Section~\ref{sub:examplec}}. In particular, the Stokes~$V$ zero-crossing point is shifted to the red by 7.75~km~s$^{-1}$.

In addition, we have examined the shape of the Stokes~$V$ profiles within patches harboring supersonic downflows. The top panel of Figure~\ref{fig:parche_stokesv} shows a close-up of the region labeled a in Figure~\ref{fig:intensity_observables} where a large downflowing patch is observed. It has an elliptical shape, with a length of about 2\arcsec\, and a width of 0\farcs7. The Stokes~$V$ zero crossing points yield LOS velocities of order 9~km~s$^{-1}$ in some places of the patch. We find that not all the Stokes~$V$ profiles in this patch show two lobes, they usually have three lobes, similar to the ones described by Franz \& Schlichenmaier (\citeyearads{2009A&A...508.1453F}, \citeyearads{2013A&A...550A..97F}). 

In fact, we observe that the Stokes~$V$ profiles within the patch show a smooth transition from regular penumbral profiles (two unshifted lobes with the same polarity as the sunspot) just outside the patch to supersonic ones (two reversed and strongly redshifted lobes) in its center. This can be seen in the lower panels of Figure~\ref{fig:parche_stokesv}, where we display the Stokes~$V$ profiles emerging along a cut between points A and B (white line in the top panel). The upper Stokes~$V$ profile corresponds to a pixel embedded in a normal penumbral region. It has two lobes and is almost unshifted. The second Stokes~$V$ profile shows three lobes. The third lobe is redshifted and small. In the following panel, this lobe has grown and its amplitude is comparable to that of the first lobe, which decreases until it vanishes. When this occurs (middle panel), the Stokes~$V$ is regular and reversed. This profile is emerging in the center of the patch. The next panels show the inverse situation: a small unshifted lobe grows in the blue continuum and becomes greater than the third one that disappears in the penumbra at point B (three last panels, respectively). 

Therefore, Figure~\ref{fig:parche_stokesv} suggests that regions close to the edges of the patch are occupied by a mixture of a regular penumbral component and another supersonic component, i.e., two Stokes~$V$ profiles with opposite polarity and different Doppler shifts, which is observed as three-lobed profiles. However, central areas show only a supersonic component (middle panel): reversed and strongly redshifted Stokes~$V$ profiles. Although this patch shows a well-organized spatial distribution, it does not occur in all cases.

\subsection{Inversion of supersonic profiles}
\label{sub:inversions}

We performed an inversion of the Stokes profiles obseved in the ten patches with the SIR code \citepads{1992ApJ...398..375R} to derive the magnetic field vector and the LOS velocity. Since we are interested in pixels with redshifted LOS velocities in the patches, we have considered only those pixels with Stokes~$V$ zero-crossing shifts larger than 5~pm (i.e., $\sim$2.5~km~s$^{-1}$).

\begin{figure}[!t]
\centering
\includegraphics[trim = {0cm 1.5cm 2cm 3cm}, clip, width = 0.50\textwidth]{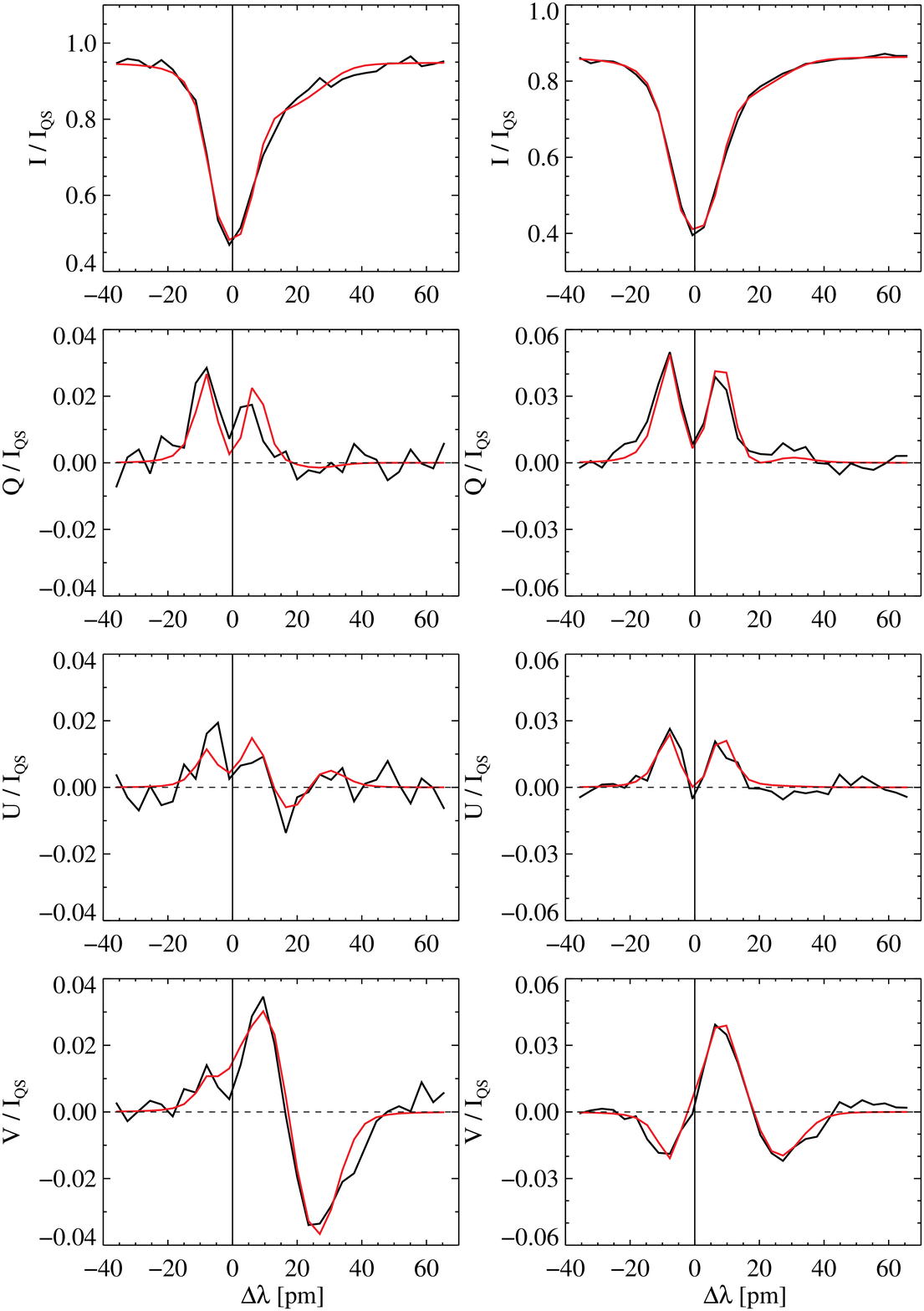} 
\caption{Stokes profiles emerging from a supersonic Evershed patch and best fit achieved by a two component model atmosphere (black and red, respectively). The four Stokes profiles in the left column correspond to Example~1 of Figure~\ref{fig:ejemplos_generales} and those in the right one emerged from a pixel of the same patch.}
\label{fig:ajustes_inversiones}
\end{figure}

\begin{table}[!t]
\centering
\begin{tabular}{ccc}
\textbf{Parameter} & \textbf{Fast component} & \textbf{Slow component} \\ 
\hline \hline
LOS velocity & 11~km~s$^{-1}$ & 0.5~km~s$^{-1}$ \\ 
Field strength & 2~kG & 1~kG \\ 
Field inclination & 50\degree & 130\degree \\ 
Filling factor & 95\% & 5\% \\ 
\hline \\
\end{tabular} 
\caption{Initial atmospheric parameters considered for the inversions.}
\label{tab:modelos_iniciales}
\end{table} 

\begin{figure}[!t]
\centering
\includegraphics[trim = {0.75cm 1cm 2cm 2.9cm}, clip, width = 0.50\textwidth]{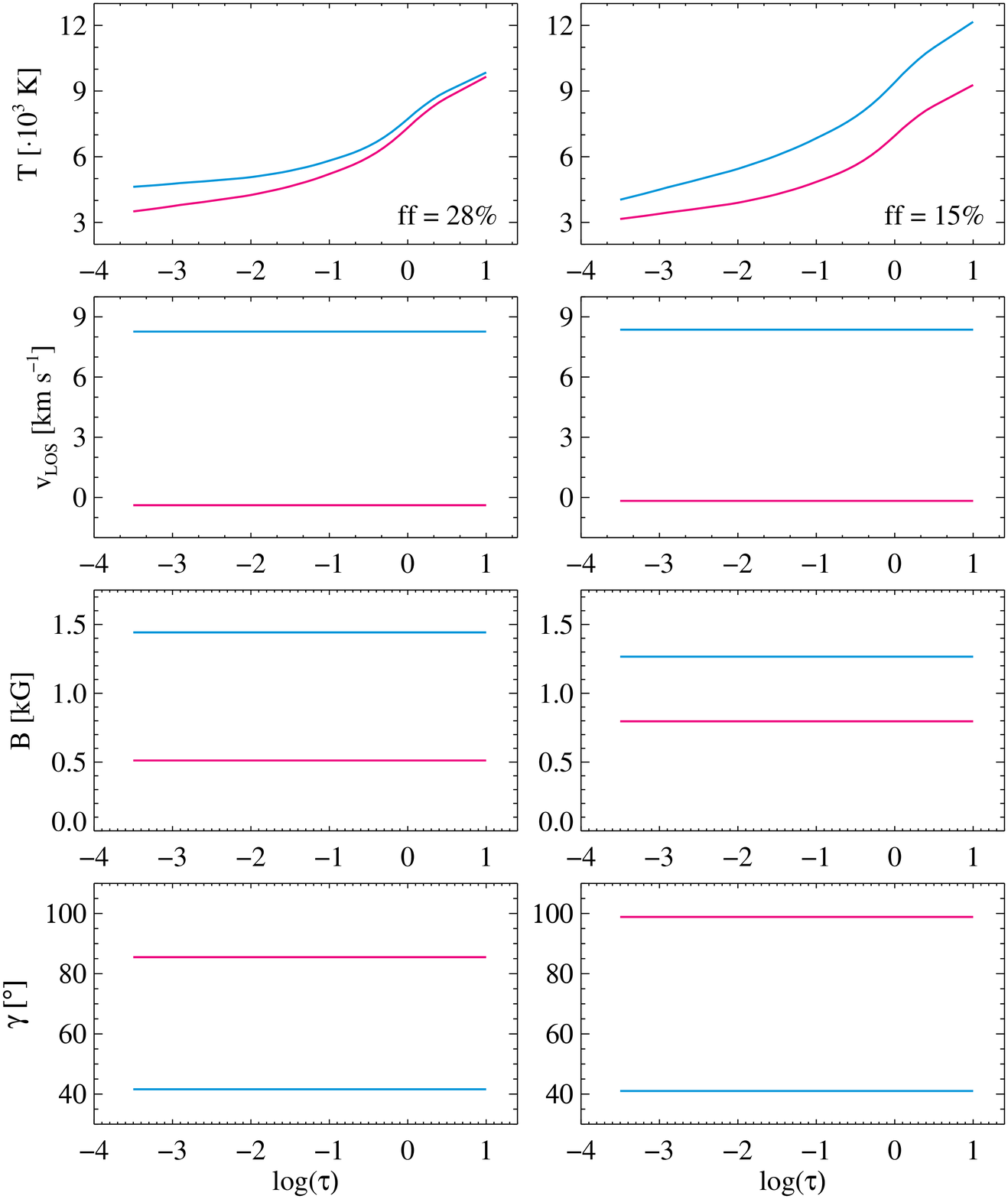} 
\caption{Atmospheric parameters resulting from the inversion of the Stokes profiles shown in Figure~\ref{fig:ajustes_inversiones}. The fast and slow components are represented in blue and magenta lines, respectively.}
\label{fig:modelos_inversiones}
\end{figure}

The Stokes profiles in these patches show a peculiar behavior: Stokes~$I$, $Q$, or $U$ do not reveal indications of supersonic velocities, except for an extended red wing in Stokes~$I$. The strong redshift is very obvious in Stokes~$V$. Linear polarization is also strong, regardless of the number of lobes in Stokes~$V$.  Figure~\ref{fig:ajustes_inversiones} illustrates this behavior for two pixels harboring supersonic downflows. This suggests that two very different atmospheres coexist in the pixel (or two layers with different physical properties along the LOS): one with fast flows and rather vertical fields of opposite polarity, accounting for the redshifted and reversed Stokes~$V$ signal and another having slow or no flows and a more horizontal field of the same polarity as the sunspot, producing the unshifted linear polarization signals. 

Describing these profiles with a single depth-stratified atmosphere is challenging, and requires strong vertical gradients in physical parameters. We have chosen a simpler approach that allows us to reproduce the profiles using two components, each of them with a constant velocity and magnetic field vector. This approach does not represent two unresolved components within the same pixel, because in fact the patches are well extended and spatially resolved, but it allows to account for two different regimes in the atmosphere and is warranted by the results in Figures~\ref{fig:parche_stokesv} and~\ref{fig:ajustes_inversiones}.

We have initialized our inversions with two components that capture the essence of this situation: the slow component is assumed to be almost at rest and has the same polarity as the spot, whereas the fast component is initialized with a strong downflow and opposite magnetic polarity. We have set the field inclination to values of 50\degree\, and 130\degree, respectively, to ensure response in all Stokes parameters to changes in the magnetic field vector.
The initial parameters used for each component are shown in Table~\ref{tab:modelos_iniciales}. During the inversion, the LOS velocity and the magnetic field vector of both were assumed to be constant with height. The temperature was allowed to have three nodes. Macroturbulence, microturbulence and stray light were set to zero. The two components are mixed according to a filling factor.

Although we are using a relatively simple two-component model, our best fit profiles show excellent agreement with the observations (see Figure~\ref{fig:ajustes_inversiones}). They capture the shapes of the observed profiles within the noise, including the three lobes of Stokes~$V$. Figure~\ref{fig:modelos_inversiones} shows the atmospheric parameters resulting from the inversion. In both pixels, the temperature monotonically decreases with height and the fast atmosphere is always hotter than the slow one at all heights. The LOS velocity of the fast atmosphere is supersonic ($\sim$8.5~km~s$^{-1}$), while that of the slow one is zero or slightly blueshifted. The magnetic field inferred for both components represents different conditions. The fast atmosphere harbors a magnetic field strength of about 1.3~kG with a rather vertical inclination of 40\degree\,(its polarity is opposite to that of the sunspot\footnote{The polarity of the sunspot is negative. Therefore, inclinations between 90\degree\,and 180\degree\,are of the same polarity as the spot, while those ranging from 0\degree\,to 90\degree\,indicate reversed polarity.}), and the slow one is weaker ($\sim$0.5~kG) and nearly horizontal, of the same or opposite polarity.

\section{Results}
\label{sec:results}

\begin{figure}[!t]
\centering
\includegraphics[trim = {0.6cm 19.25cm 2cm 2.75cm}, clip, width = 0.52\textwidth] {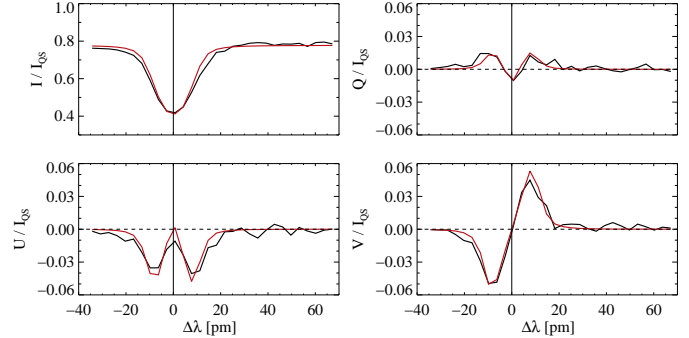} 
\caption{Regular Stokes profiles emerging from the penumbra and best fits profiles given by the one-component inversion (plotted in black and red, respectively).}
\label{fig:ajuste_1_comp}
\end{figure}

\begin{figure*}[!t]
\centering
\includegraphics[trim = {9.25cm 6cm 2.5cm 2.25cm}, clip, height = 0.72\textheight]{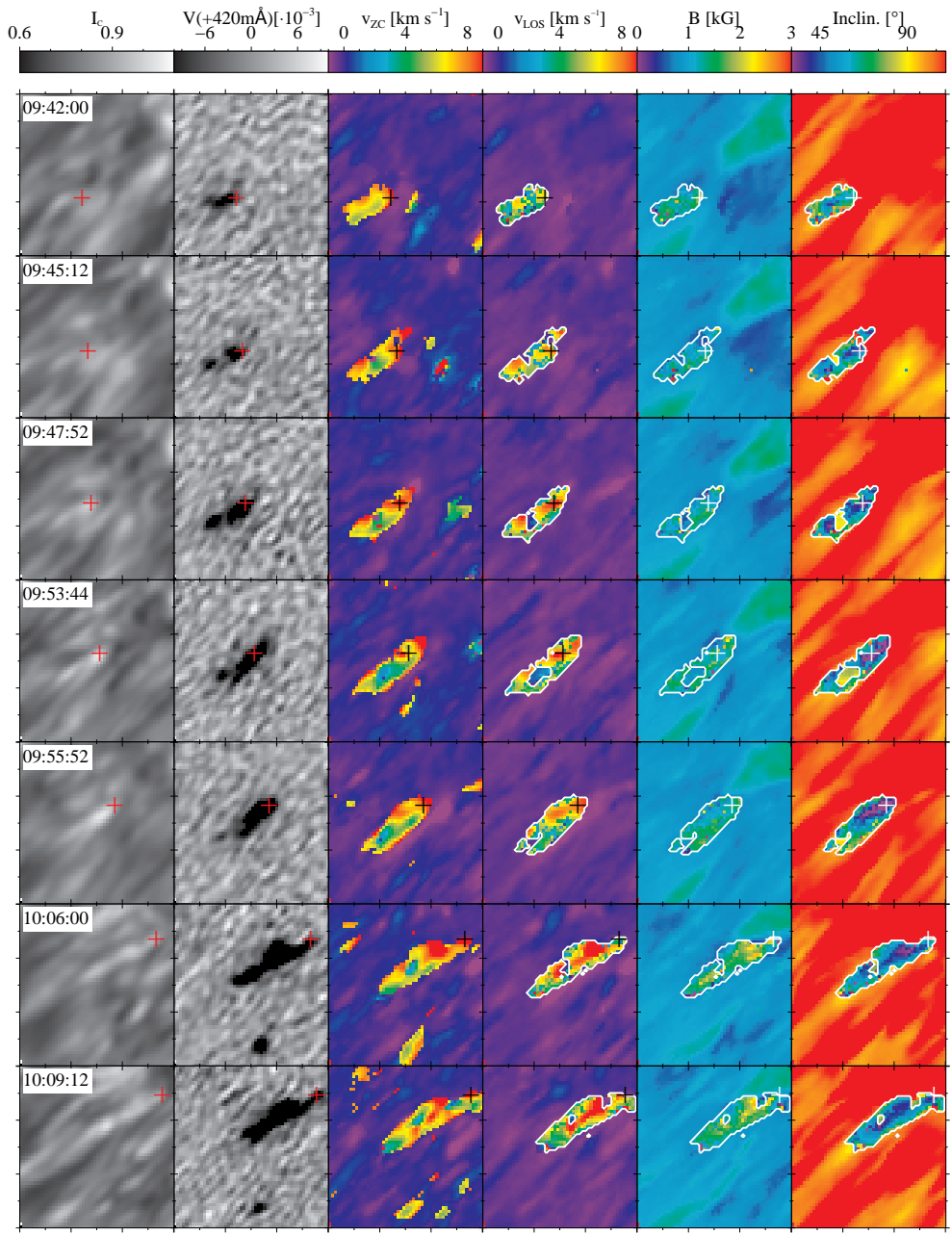} 
\caption{Temporal evolution of the physical parameters of Example a. From left to right: continuum intensity filtered for subsonic oscillations, magnetogram, LOS velocity given by the Stokes~$V$ zero-crossing wavelength, and LOS velocity, magnetic field strength and inclination of the fast atmosphere retrieved from the two-component inversions. In the last three rows, the results from one-component inversions give information about the surroundings of each patch (enclosed by white contours). A plus symbol indicates the tracked position. Time increases from top to bottom, but the panels are not equally spaced. Each major tickmark represents 1\arcsec.}
\label{fig:ejemplo1}
\end{figure*}

In this section we present the temporal evolution of the patches with supersonic Evershed downflows. Their main characteristics are described by three examples located in different penumbral positions (labeled as a, b and c in Figure~\ref{fig:intensity_observables}). We also show the physical properties obtained from a statistical analysis of all the pixels having supersonic velocity within the patches. Finally, we compare them with the properties of their surroundings.

\begin{figure*}[!t]
\centering
\includegraphics[trim = {9.25cm 6.75cm 2.25cm 3cm}, clip, height = 0.77\textheight]{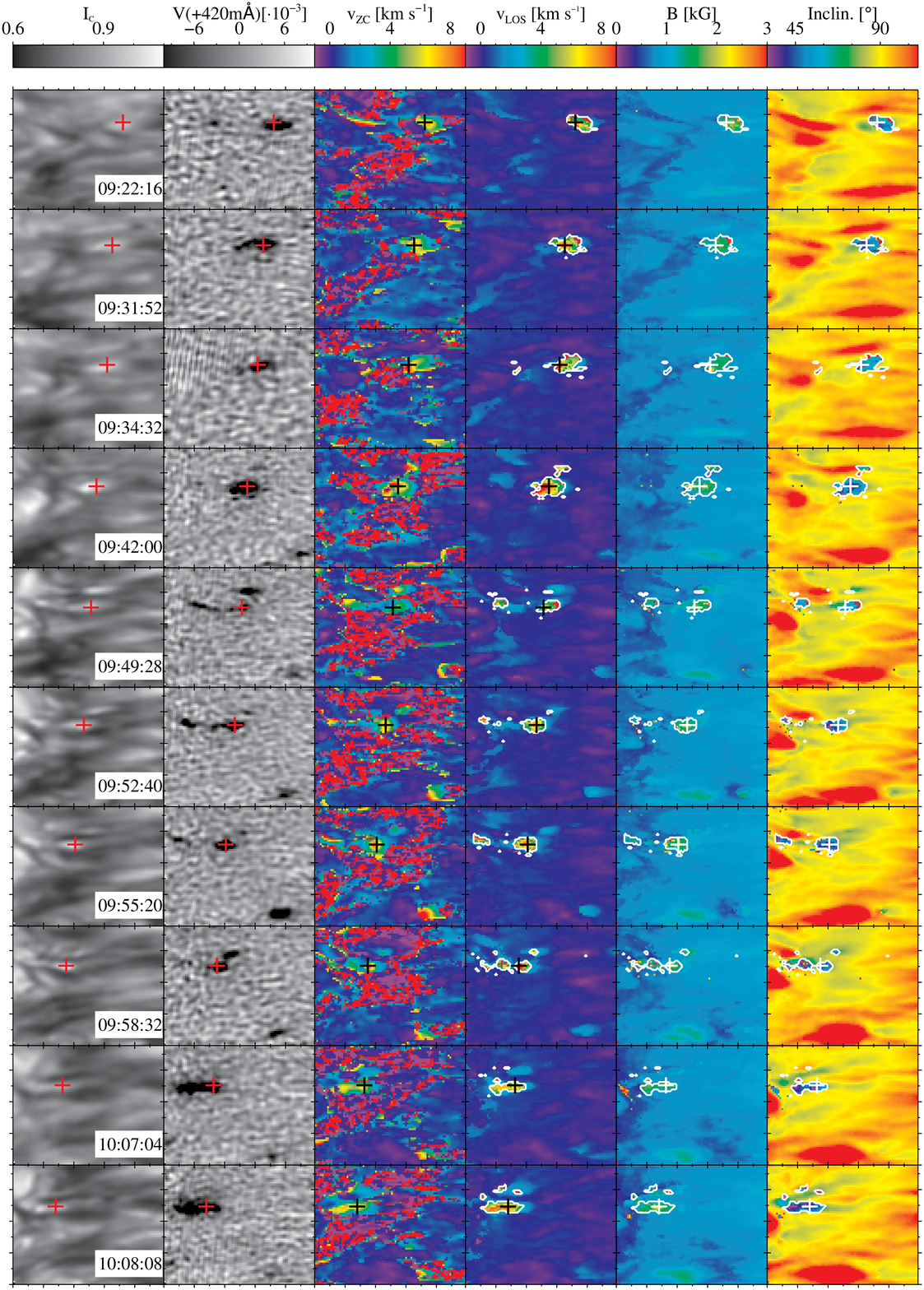} 
\caption{Temporal evolution of the physical parameters of Example b. The layout is the same as in Figure~\ref{fig:ejemplo1}.}
\label{fig:ejemplo2}
\end{figure*}

\subsection{Temporal evolution of supersonic Evershed downflows}
\label{sub:temporal}

We study the temporal evolution of the physical parameters and Stokes profiles of each patch by selecting a position within the patch and tracking it manually during the lifetime of the patch. To choose that position, we considered the maximum LOS velocities computed by the Stokes~$V$ zero-crossing points, but we also double-checked our selection with the LOS velocity inferred for the fast component. In addition, we required continuous variations of the position of the selected pixel along its lifetime to obtain smooth trajectories.

\begin{figure}[!t]
\centering
\includegraphics[trim = {1cm 3.5cm 0cm 6cm}, clip, width = 0.4\textwidth]{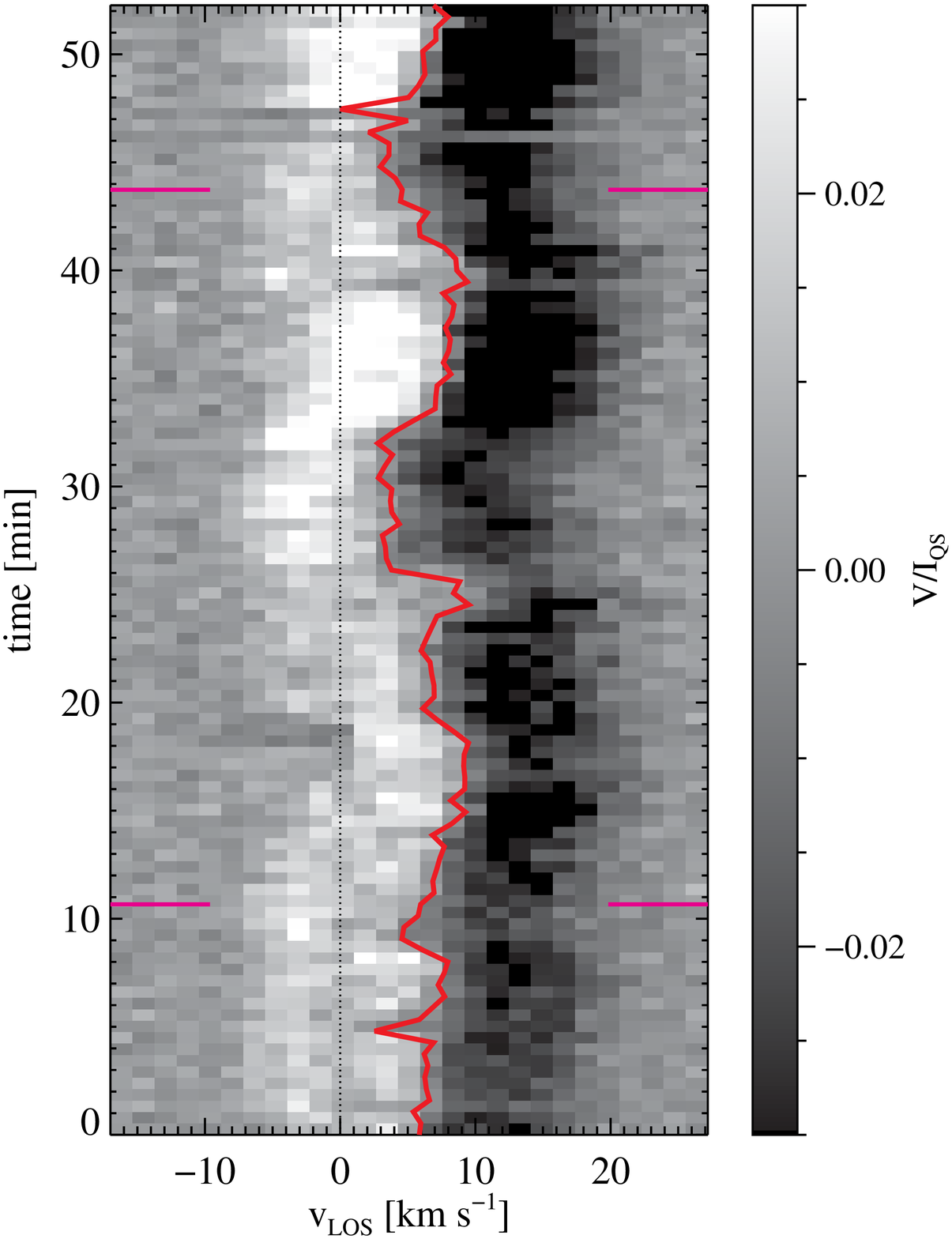} 
\caption{Temporal evolution of the Stokes~$V$ profiles observed in the tracked position of Example b during all the sequence. Time goes from bottom to top with a sampling of 32~s. The red line shows the Stokes~$V$ zero-crossing points of the profiles. The dotted vertical line represents v$_{LOS}$~=~0. The horizontal dashes delimit the time sequence shown in Figure~\ref{fig:perfiles_ejemplob}.}
\label{fig:spectravej2}
\end{figure}

\begin{figure}[!t]
\centering
\includegraphics[trim = {1.75cm 0cm 4cm 1.25cm}, clip, width = 0.44\textwidth]{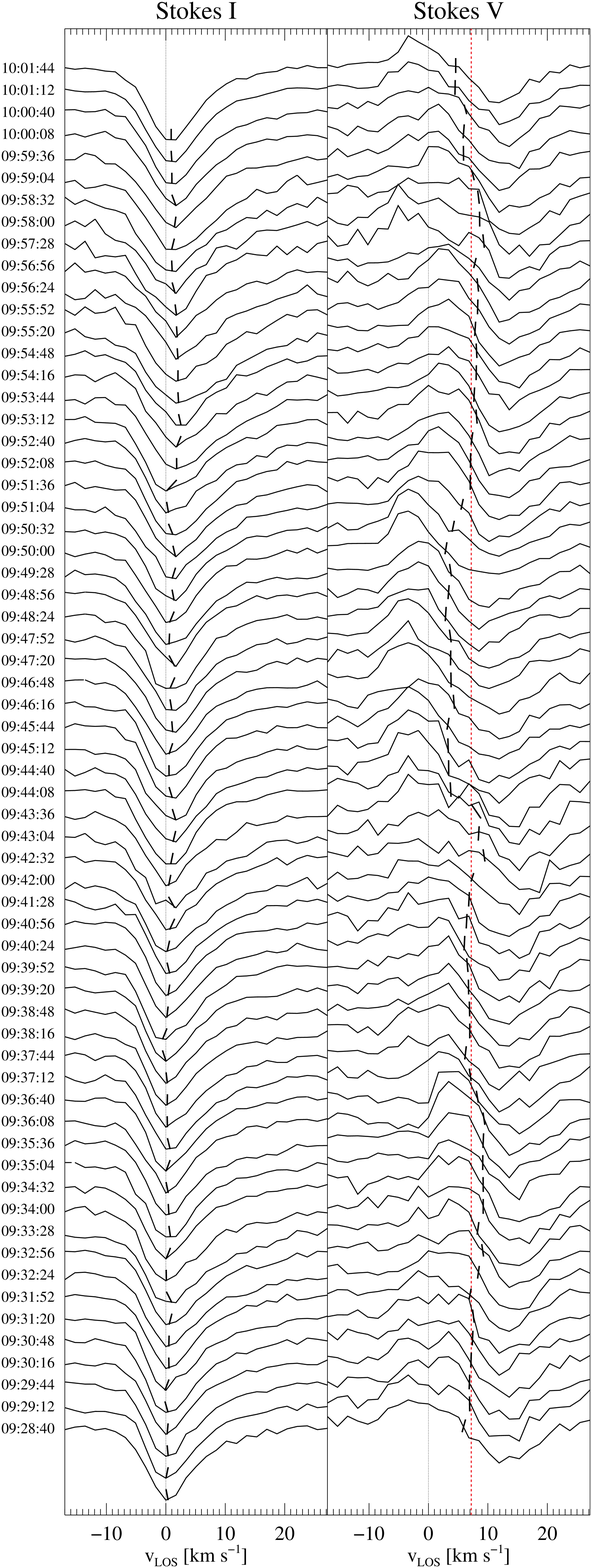} 
\caption{Temporal sequence of the Stokes~$I$ (left) and~$V$ (right) profiles observed in the tracked position of Example b between 09:28:40~UT and 10:01:44~UT (marked with horizontal dashes in Figure~\ref{fig:spectravej2}). The dashed lines connect the minima of the intensity profiles and the Stokes~$V$ zero-crossing wavelengths. The red dotted vertical line indicates the sound speed in the photosphere. Black dotted vertical lines show the rest position. The profiles are arbitrarily shifted in the vertical direction. Time is shown on the left side and the temporal sampling is 32~s.}
\label{fig:perfiles_ejemplob}
\end{figure}

We performed one-component inversions using SIR (specifically, the parallel implementation of \citeads{2015IAUS..305..251T}) to derive the conditions in the surroundings of the patch, since the two-component inversions were carried out only in pixels with zero-crossing velocities larger than $+$2.5~km~s$^{-1}$. For the one-component inversions, we assumed height-independent stratifications of magnetic field strength, inclination, azimuth and LOS velocity, and three nodes for temperature. The macroturbulence, microturbulence, and stray light were set to zero. Despite the simplicity of this model, Figure~\ref{fig:ajuste_1_comp} illustrates a good fit to the observed penumbral profiles.

The complete sequences for these examples are available in the electronic Journal. The movies also show the Stokes profiles and the values of the atmospheric parameters of the selected pixel during the tracking. For reference, the average penumbral profile in a 3\arcsec\, x 3\arcsec\, box surrounding the pixel of interest is plotted with thin solid line in the Stokes~$V$ panel.

\begin{figure*}[!t]
\centering
\includegraphics[trim = {8.25cm 6cm 2cm 2.75cm}, clip, width = 0.72\textwidth]{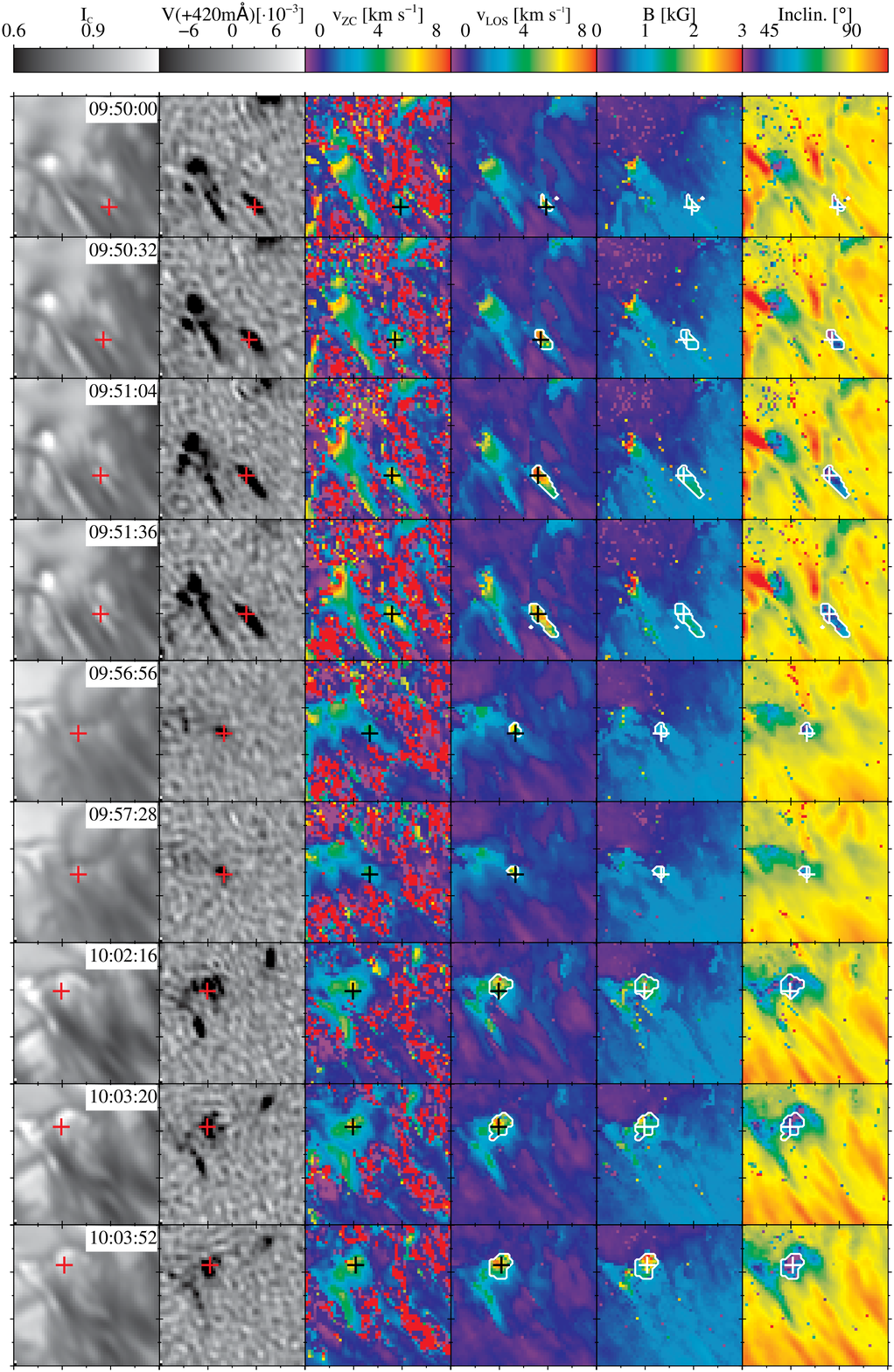} 
\caption{Temporal evolution of the physical parameters of Example c. The layout is the same as in Figure~\ref{fig:ejemplo1}.}
\label{fig:ejemplo3}
\end{figure*}

\subsubsection{Example a}
\label{sub:examplea}

Figure~\ref{fig:ejemplo1} shows the temporal evolution of a patch with supersonic flows in the mid center side penumbra. The patch is located at the end of a penumbral filament, specifically in a bright feature that resembles a penumbral grain moving outwards. For visualization purposes, only the most significant non-consecutive scans are displayed. The patch is clearly observed in all maps from 09:42:00~UT. Usually, the Stokes~$V$ profiles observed in the tracked position have three lobes, although sometimes they are regular and strongly redshifted, as can be seen in the movie corresponding to this example. Stokes~$I$, $Q$ and $U$ do not reveal clues of supersonic behavior.

The tracked position (marked with a plus symbol in each row of Figure~\ref{fig:ejemplo1}) is usually at the boundaries of the patch and of a bright feature. In the second row, there is a merging with another bright structure, which is also identified in the magnetogram, the LOS velocity and the inclination panels. During its lifetime, this patch merges with two more patches containing supersonic downflows, which increases the LOS velocity of the tracked position. From the second to the sixth row, the selected pixel shows a supersonic downflow, from 7.2~to~9~km~s$^{-1}$, concentrated on the downwind side of the bright grain that is moving outwards. This trend ends at 10:06:00~UT when the patch is fragmented and the LOS velocity is much lower (3~km~s~$^{-1}$ in the fast atmosphere). However, it sets in a few scans later, becoming supersonic again. Thus, the patch is supersonic during four time intervals, which last between 2~and 8~minutes (visible in 4~or 16~frames, respectively).

The magnetic field remains rather vertical with inclinations between 25\degree\,and 55\degree\, respect to the vertical. The field strength usually ranges between 1~and~1.5~kG, which are typical penumbral values. Interestingly, the more vertical inclinations are related to stronger magnetic fields only during interactions.

\subsubsection{Example b}

This case consists of a patch located between the mid and outer limb side penumbra and is represented in Figure~\ref{fig:ejemplo2}. Despite our time sequence does not cover its full lifetime, it might end beyond the outer sunspot border judging by its trajectory. It remains visible during all the sequence, i.e., $\sim$50~minutes. It is located in a very bright structure ($\sim$1.1~I$_{QS}$) that moves outward at the outer end of a penumbral filament. 

The fast atmosphere usually achieves LOS velocities greater than 5~km~s$^{-1}$, even reaching more than 9~km~s$^{-1}$ at times. Despite some fluctuations, the LOS velocity of the fast atmosphere is recurrent and becomes supersonic at least five times, with durations between 1 and 6~minutes. Some of these recurrences can be observed in Figure~\ref{fig:ejemplo2}. The LOS velocity shows supersonic velocities in all rows (except in the fifth and the ninth, which are used to illustrate the situation when the LOS velocity is not supersonic).

In addition, two fragmentations occur. They start as brightenings that move towards the head of the patch and induce a horizontal acceleration that elongates the patches observed in the continuum intensity and in the LOS velocity maps (fourth row). Although we detect them as a single structure in intensity, it breaks in the LOS velocity maps. Then, the LOS velocity of the separated region grows as well as its intensity, fading two or three frames later. Simultaneously, the fragment of the patch lagging behind continues its trajectory without being influenced by the fragmentation.

The magnetic strength covers a wide range of values, from 1~to 2~kG, and its inclination oscillates between 20\degree\,and 60\degree, pointing to rather vertical magnetic field lines with reversed polarity. As in the previous example, the more vertical fields are found in regions of stronger magnetic field.

Figure~\ref{fig:spectravej2} displays the temporal evolution of the circular polarization signal emerging from the tracked position. The Stokes~$V$ profiles are regular and redshifted. The red line represents the LOS velocity obtained from the Stokes~$V$ zero-crossing points. It reaches high values, sometimes more than 9~km~s$^{-1}$. Figure~\ref{fig:perfiles_ejemplob} shows in detail the Stokes~$I$ and~$V$ profiles observed during the time interval between the pink horizontal dashes plotted in Figure~\ref{fig:spectravej2}. During this period, the shift of the Stokes~$V$ profiles changes smoothly and two recurrences of supersonic speeds can be observed at 09:42:00 and 09:53:44~UT.

\subsubsection{Example c}
\label{sub:examplec}

Figure~\ref{fig:ejemplo3} shows Example~c, which is located in the mid region of the limb side penumbra. Here, we have followed the leading patch moving outwards after a fragmentation. This patch lasts 12 minutes, perhaps it lives longer but we are not able to identify it in more frames. It is firstly detected as a bright structure at the outer end of a filament. Four frames later it becomes brighter and it is accelerated horizontally, showing a strongly redshifted Stokes~$V$ profile. The pixel that we used in the tracking is centered in the patch in all parameters except in the magnetogram, where it turns out to be slightly shifted to the left.

The LOS velocity of the fast atmosphere suddenly increases from $\sim$4~to~8~km~s$^{-1}$ (two first rows of Figure~\ref{fig:ejemplo3}) and, subsequently, the patch fragments. In this case, the portion of the patch with enhanced LOS velocity is motionless and darker and fades nine frames later. In the meantime, the leading portion gets brighter and accelerates toward the outer penumbral border, but its LOS velocity decays. When the patch reaches the moat region around the sunspot, the LOS velocity abruptly grows from 6~to~8~km~s$^{-1}$ (seventh and eighth rows) and falls down afterwards. In addition, there is a large increase of magnetic strength (by $\sim$1~kG) when the tracked position protrudes into the moat region, causing broader Stokes~$V$ profiles. At that moment, the maximum magnetic field strength is of order 2.5~kG. Similarly to the previous examples, pixels with supersonic LOS velocities show reversed polarity during all the sequence. This case shows only two brief recurrences, with a duration of 1~minute each one.

\subsection{Statistical analysis}

We have characterized the main properties of the supersonic Evershed downflows by carrying out a statistical analysis of the parameters retrieved for the fast atmosphere. We considered the best fits of all our inversions (except in pixels with three-lobed Stokes~$V$ profiles with similar lobe amplitudes, where the inversions were not as successful as in other cases). In addition, we used the $\chi^{2}$-values of the fit to decide if the inversion was good enough to be considered.

This statistical analysis is based on 16486 pixels belonging to the ten supersonic patches located in the mid and outer penumbra and spanning their whole lifetime. A total of 3750 pixels show supersonic LOS velocities in the fast atmosphere. In addition, we compared their physical parameters with the penumbral surroundings. For this purpose, we considered a 3\arcsec\, x 3\arcsec\, area surrounding each patch where we performed one-component inversions as explained in Section~\ref{sec:results}. This reference set consists of 1.3$\cdot$10$^{6}$ pixels.

\begin{figure}[!t]
\centering
\includegraphics[trim = {1.1cm 0.5cm 4.5cm 0.5cm}, clip, width = 0.52\textwidth] {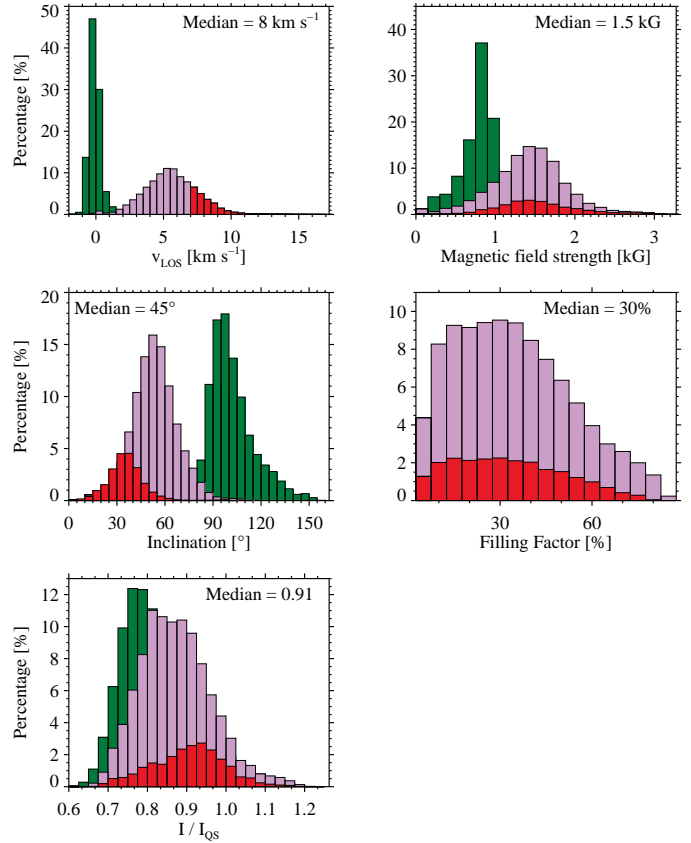} 
\caption{Histograms of LOS velocity, magnetic field strength, inclination and filling factor of the fast atmosphere, plus continuum intensity. Purple, red and green distributions represent all pixels inside the patch, those having supersonic LOS velocities, and the background, respectively. The median value of the red distributions is indicated in the upper corner of the panels.}
\label{fig:histo}
\end{figure}

Figure~\ref{fig:histo} displays histograms of the LOS velocity, field strength, inclination, and filling factor of the fast atmosphere resulting from inversions, and continuum intensity. The purple distributions are obtained considering all the pixels within the patches. Among them, the red ones represent to pixels with supersonic velocities. Both distributions are normalized to the total number of pixels contained in the patches. The green distributions correspond to those from the surrounding penumbral regions and are normalized to the amount of pixels of this sample.

The distributions displayed in Figure~\ref{fig:histo} differ significantly between groups, mainly in the LOS velocity. Pixels within the patches show a broad range of positive LOS velocities, between 2 and 10~km~s$^{-1}$, with a peak around 5.5~km~s$^{-1}$. Most of the supersonic pixels have LOS velocities between 7.5 and 9.5~km~s$^{-1}$, although in some cases they even reach 15~km~s$^{-1}$. The supersonic subsample has a mean value of 8~km~s$^{-1}$. In contrast, the surrounding penumbral regions show a moderate velocity dispersion around zero up to $\pm$1.5~km~s$^{-1}$. The histogram of magnetic field strength in the supersonic patches is similar to those obtained from all the pixels within the patches, where most values are around 0.5 and 2~kG. The former distribution shows a median value of 1.5~kG. In the surrounding penumbral regions, the field strengths vary between 0.2 and 1~kG. The distributions of the filling factors in the supersonic and those inside the patches are akin and the supersonic pixels have a median of 30\%. Here, the background distribution is not represented, since its physical parameters have been retrieved from one-component inversions and the filling factor is always 100\%.

The difference of the field inclination distributions obtained for each group is also significant. The inclination in the supersonic pixels varies between 5\degree\,and 65\degree\,with a median of 45\degree, while it reaches 110\degree\,when we consider all pixels within the patches. Therefore, the magnetic field of the supersonic pixels is more vertical with opposite polarity than that of the sunspot. The background usually shows the same polarity of the sunspot and in some cases, it is practically horizontal. The distribution of continuum intensity of the supersonic pixels is shifted to higher values compared with the other sets. This demonstrates that the supersonic Evershed downflows tend to coincide with bright features, as we inferred from their temporal evolution. The median continuum intensity in the supersonic pixels is 0.91.

In Table~\ref{tab:mediana} we summarize the median values of the distributions for the three populations. Here we can see that supersonic Evershed downflows are observed as brighter locations where magnetic field lines are more vertical.  

\begin{table}[!t]
\centering
\begin{tabular}{cccc}
\textbf{Physical} & \textbf{Supersonic} & \textbf{Pixels within} & \textbf{Regular} \\
\textbf{parameter} & \textbf{pixels} & \textbf{patches} & \textbf{penumbra} \\
\hline \hline
v$_{LOS}$ [km~s$^{-1}$] & 8.0 & 5.5 & $-$0.2 \\ 
B [kG] & 1.5 & 1.5 & 0.8 \\ 
$\gamma$ [\degree] & 45 & 55 & 100 \\ 
Fill. factor & 0.30 & 0.33 & 1.00 \\ 
I$_{cont}$ & 0.91 & 0.82 & 0.80 \\
\hline \\
\end{tabular} 
\caption{Median values of the distributions shown in Figure~\ref{fig:histo}.}
\label{tab:mediana}
\end{table}

\section{Discussion}
\label{sec:discussion}

Our analysis provides detailed information on the physical parameters of the supersonic Evershed flows discovered spectroscopically and characterized so far using $\sim$1\arcsec\,resolution spectropolarimetric data. In this section we describe how our results expand what is known about these flows.

\subsection{Relation between detected patches and bright structures}

The temporal evolution of the patches reveals that these are co-spatial with bright, seed-like intensity features (see Section~\ref{sub:temporal}). These bright structures resemble penumbral grains that move outwards, first reported by \citetads{1992SoPh..140...41W}. Later, \citetads{1999A&A...348..621S} found a dividing line in the penumbra, approximately located in the mid region, where most penumbral grains inside this line move towards the umbra while those outside migrate to the quiet Sun, which was also corroborated by \citetads{2001A&A...380..714S}. Supersonic Evershed downflows are found also in the mid and outer penumbra.

\citetads{2007PASJ...59S.593I} found that the regions with enhanced Stokes~$V$ signal in the blue wing correspond to bright inner penumbral grains, being the sources of the Evershed flow (e.g.,~\citeads{2006ApJ...646..593R}). But they did not mention any relation between outward moving penumbral grains and enhanced Stokes~$V$ signal in the red wing. However, their Figure~5 shows a relation between enhanced Stokes~$V$ signal in the red wing and a bright position. This is also found in \citetads{2010ASSP...19..193B}. In our study, Figures~\ref{fig:ejemplo1}, \ref{fig:ejemplo2} and ~\ref{fig:ejemplo3} also show a connection between enhanced signals in the far red wing magnetogram and bright intensity structures. 

Finally, the temperature stratifications obtained from our two-component inversions (Figures~\ref{fig:modelos_inversiones} and~\ref{fig:ajustefinal_irreal}) unveil a monotonic decrease with height. Therefore, the bright intensity features we observe could be produced by temperature enhancements caused by a sudden halt of the supersonic Evershed downflows at lower photospheric layers, which are denser and therefore represent an obstacle for the downward flow. 

\subsection{Comparison with previous studies}

Our results confirm quantitavely the existence of supersonic Evershed downflow in a sunspot penumbra, as previously suggested by other authors (Section~\ref{sec:introduction}). Our LOS velocities are similar to those observed by \citet{bumba60}, \citetads{1960IAUS...12..403S} and \citetads{2010ASSP...19..193B}, and greater than those reported by, e.g., \citetads{2004A&A...427..319B} and \citetads{2007PASJ...59S.593I}. We find differences with the results of \citetads{2001ApJ...549L.139D}, who detected Evershed supersonic downflows concentrated in cold magnetic tubes with LOS velocities up to 16~km~s$^{-1}$. In our spot, supersonic Evershed downflows tend to be bright (hot) and reach LOS velocities of 16~km~s$^{-1}$ in rare occasions, representing only 1\%\, of the observed supersonic pixels. 

Our LOS velocities are comparable to those inferred by \citetads{2013A&A...557A..24V} in a sunspot located close to the disk center. However, we retrieve magnetic field strengths between 1~and 2~kG (only slightly larger than typical penumbral values), which are weaker than the 2.5~--~4~kG obtained by \citetads{2013A&A...557A..24V}. In addition, their downflow areas usually are of order 0.1~arcsec$^{2}$. Our smallest patches have an area of 0.3~arcsec$^{2}$ (tiny patches in Figure~\ref{fig:ejemplo2}) and can reach up to 2.5~arcsec$^{2}$.

Another significant difference with the results of \citetads{2013A&A...557A..24V} is the location of supersonic downflows. They usually found them in the outer boundary of the spot. Specifically, they reported supersonic LOS velocities at the end of complex features, being weaker in simple filaments. Here, we detect supersonic Evershed downflows at the end of both simple and complex filamentary features between the mid and outer penumbral regions (see Figure~\ref{fig:distribucion}), having large LOS velocities regardless of their location in the spot. 

We find that patches harboring supersonic Evershed downflows are very often observed as bright features moving outward in intensity filtergrams. Furthermore, temperatures inferred for the fast component from inversions rise significantly with depth. Both clues point to the existence of a shock in deeper layers. However, \citetads{2013A&A...557A..24V} did not observe significant temperature enhancements in the downflows.

\begin{figure}[!t]
\centering
\includegraphics[trim = {0.25cm 1cm 1cm 2.5cm}, clip, width = 0.525\textwidth] {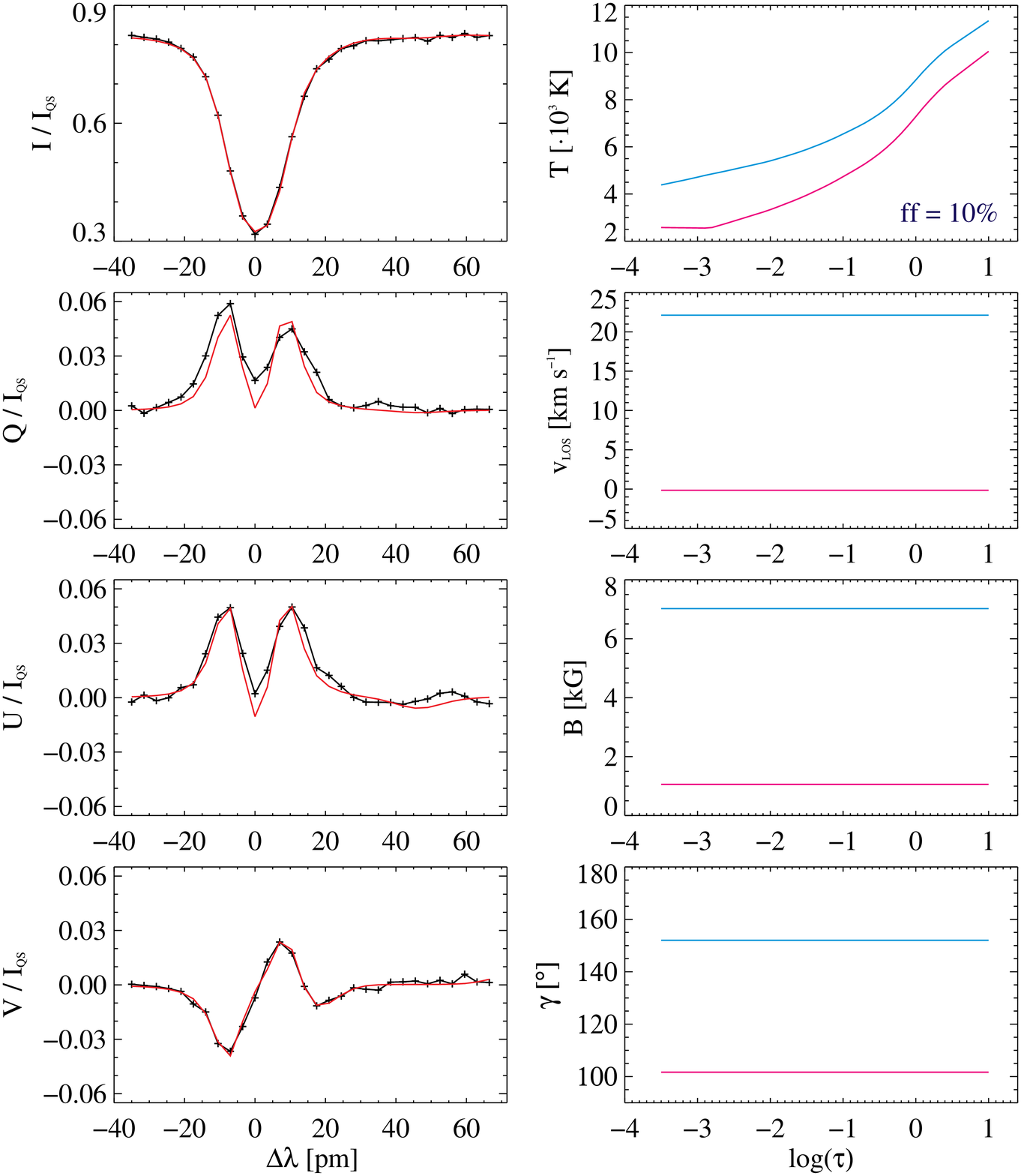} 
\caption{Right panels: Observed and best fit Stokes profiles (black and red, respectively). Left panels: Atmospheric parameters inferred for the fast (blue) and the slow components (magenta) from the inversion. The filling factor of the fast component is shown in the lower left corner of the upper panel.}
\label{fig:ajustebueno_irreal}
\end{figure}

\begin{figure}[!t]
\centering
\includegraphics[trim = {0.25cm 19.25cm 1cm 3cm}, clip, width = 0.525\textwidth] {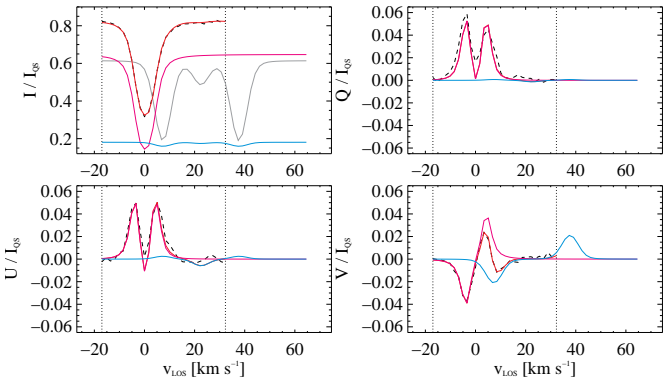} 
\caption{Synthetic profiles resulting from the atmospheric stratifications of the fast and the slow atmospheres (blue and magenta lines, respectively) plotted in Figure~\ref{fig:ajustebueno_irreal}. The grey line shows the blue one multiplied by a scale factor for better visibility. As reference, the observed and the best fit profiles are also shown here in black and red, respectively. The dotted vertical lines delimit the observed spectral range.}
\label{fig:sintesis_irreal}
\end{figure}

We do not find any example similar to the peculiar supersonic downflows reported by \citetads{2013A&A...557A..24V}. Despite we retrieve also high LOS velocities and magnetic field strengths for the fast component in some pixels (see Figure~\ref{fig:ajustebueno_irreal}), we cannot trust them since these very large parameters produce synthetic profiles with a significant portion being out of the observed spectral range (Figure~\ref{fig:sintesis_irreal}). The inversion code combined a strongly shifted two-lobed Stokes~$V$ profile with an unshifted one to recreate the observed three-lobed Stokes~$V$ profile, without any information about how the profiles would look like outside the spectral range.

\begin{figure}[!t]
\centering
\includegraphics[trim = {0.25cm 1cm 1cm 2.5cm}, clip, width = 0.525\textwidth] {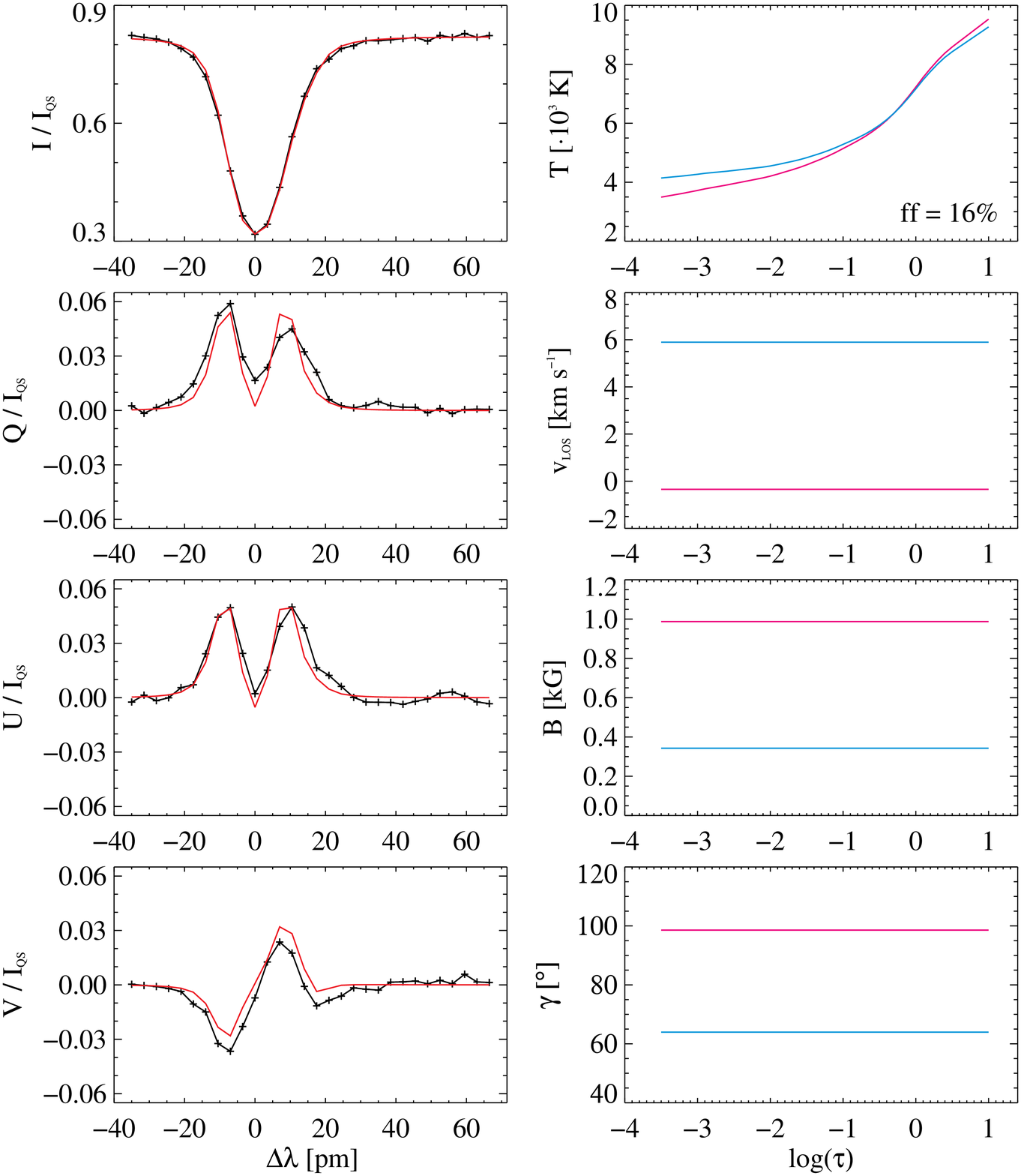} 
\caption{best fit profiles and atmospheric parameters inferred from an inversion of the same pixel as in Figure~\ref{fig:ajustebueno_irreal} considering as initial models the results from a close pixel with a similar Stokes~$V$ profile. The layout is the same as in Figure~\ref{fig:ajustebueno_irreal}.}
\label{fig:ajustefinal_irreal}
\end{figure}

To solve this situation, we decided to repeat the inversion considering as initial models the atmospheric stratifications obtained from a nearby pixel unaffected by this problem whose Stokes~$V$ profile is similar. The results are shown in Figure~\ref{fig:ajustefinal_irreal}. Although the fit is not as good as in the previous inversion, it is still reasonable. The inversion now finds a reversed, strongly shifted and an unshifted two-lobed Stokes~$V$ profile whose combination reproduces the observed spectral signatures, being a more reliable solution.

Although these structures resemble Evershed clouds (e.g.,~\citeads{1994ApJ...430..413S};~\citeads{1994A&A...290..972R};~Cabrera Solana et al.~\citeyearads{2006ApJ...649L..41C};~\citeyearads{2007A&A...475.1067C}), we do not find any indication of two velocity packets inside the patch travelling outward as a whole. In addition, \citetads{2007A&A...475.1067C} found that the two Evershed cloud packets have opposite magnetic polarities when they cross the sunspot border, what has not been found in our observations.

A number of physical mechanisms are known to be able to drive supersonic flows in the penumbra, including siphon flows \citepads{1988ApJ...333..407T} and moving magnetic tubes \citepads{2002AN....323..303S}. Both mechanisms might well explain our observations, but a detailed comparison is difficult because the theoretical results are based on idealized models that are often simplistic and do not consider the actual conditions of the penumbra (in particular, the fact that it is a stratified atmosphere with temperatures, pressures and magnetic fields varying both vertically and horizontally).

The 3D radiative MHD simulations of \citetads{2012ApJ...750...62R}, on the other hand, show supersonic downflow regions with an average velocity of 9.6~km~s$^{-1}$ at optical depth unity. The fastest downflows occur in the outer penumbra. Very interestingly, they show up as bright structures. The magnetic field at the position of the supersonic patches is strong ($\sim$2.5--3.0~kG) and of opposite polarity. Our inversions yield similar velocities, enhanced intensities, and reversed polarities, but weaker field strengths. A comparison with the simulations of \citetads{2012ApJ...750...62R} is beyond the scope of this paper, but may shed more light on the origin of the supersonic downflows and their relation with penumbral filaments.

\section{Summary and Conclusions}
\label{sec:conclusion}

In this paper we have characterized the properties of supersonic Evershed downflows and we have described their temporal evolution for the first time. This has been possible thanks to the high spatial resolution and the excellent seeing quality of our spectropolarimetric data of a spot observed at only 6.8\degree\, from disk center. To detect supersonic Evershed downflows, we have considered information from the continuum intensity filtergrams, the LOS velocities given by the Stokes~$V$ zero-crossing wavelength, and the far-wing magnetograms, together with the LOS velocity and the magnetic field vector inferred from two-component inversions of the observed Stokes profiles.

Supersonic Evershed downflows occur in the mid and outer penumbra. They are contained in compact patches that move outward and usually do not show strong far-wing Stokes~$V$ signals. They are often observed as bright and roundish features located at the outer end of a single filament or a more complex filamentary structure in intensity that resemble outward moving penumbral grains. However, sometimes it is not easy to identify the related filament in the intensity images. The lifetimes of the detected supersonic Evershed downflows vary from 1 to more than 5 minutes, showing a variable behavior.

Most of the supersonic downflows show a LOS velocity between 7.5~and 9.5~km~s$^{-1}$, with a median value of 8~km~s$^{-1}$. We have detected peak LOS velocities of about 15~km~s$^{-1}$ in rare occasions. Their magnetic fields tend to be more vertical (by~30\degree) than those in the pixels surrounding them. The polarity of the magnetic field at the position of the supersonic Evershed downflow is reversed compared to that of the sunspot. The magnetic field strength of the supersonic downflows has a median value of 1.5~kG, which is a typical value within the penumbra. Furthermore, supersonic downflows tend to be cospatial with bright pixels, the median value of the continuum intensity being 0.91. Differences on the physical parameters are more significant when comparing supersonic pixels with those of the penumbral surroundings.

From an analysis of the temporal evolution of the patches, we find that after appearing in the penumbra their LOS velocity increases and then fades. Subsequently, two behaviors are possible: the patch disappears and there are no more supersonic Evershed downflows, or the patch remains visible and a recurrent supersonic downflow is observed some frames later. During the evolution of the patch, the regions with stronger flows also harbor the more vertical magnetic field, but this does not mean that there is a one-to-one relation between the LOS velocity and the magnetic field inclination. Furthermore, we find that patches crossing the outer penumbral border experience LOS velocity and magnetic field enhancements, with no significant changes in the inclination. Once they leave the spot, such enhancements suddenly disappear and the patches fade in the quiet Sun.

In addition, supersonic patches undergo mergings and fragmentations. When a merging occurs, the LOS velocity as well as the magnetic field inclination and the continuum intensity increase. Furthermore, there is a relation between magnetic field strength and inclination during interactions. This scenario suggests that the Evershed flow associated with different filaments can return back to the surface at the same position. In the case of fragmentations, the intensity structure brightens and is accelerated outward. At that moment, the supersonic Evershed downflow heads the feature and the patch breaks.

The relation of supersonic patches with bright intensity features moving outward suggests that the nature of inner and outer penumbral grains is different. The Evershed flow returns back to the solar surface at the end of flow channels. As the downflow reaches supersonic velocities in the dense deep layers, it stops abruptly and produces a shock. Consequently, there is a temperature enhancement that increases the continuum intensity, which is observed as a bright penumbral grain moving outward. \\

\noindent \emph{Acknowledgements.} Financial support by the Knut and Alice 
Wallenberg Foundation and by the Spanish Ministerio de Econom\'{\i}a y
Competitividad through grants ESP2013-47349-C6-1-R and ESP2014-56169-C6-1-R, 
including a percentage from European FEDER funds, are gratefully acknowledged.
This paper is based on data acquired at the Swedish 1-m Solar Telescope, operated by
the Institute for Solar Physics of Stockholm University in the Spanish
Observatorio del Roque de los Muchachos of the Instituto de
Astrof\'{\i}sica de Canarias. This research has made use of NASA's
Astrophysical Data System.

\bibliographystyle{apj}
\bibliography{biblioads}

\begin{thebibliography}{}
\expandafter\ifx\csname natexlab\endcsname\relax\def\natexlab#1{#1}\fi

\bibitem[{Bellot~Rubio(2010)}]{2010ASSP...19..193B}
Bellot~Rubio, L.~R. 2010, Astrophysics and Space Science Proceedings, 19, 193

\bibitem[{Bellot~Rubio {et~al.}(2004)Bellot~Rubio, Balthasar, \&
  Collados}]{2004A&A...427..319B}
Bellot~Rubio, L.~R., Balthasar, H., \& Collados, M. 2004, \aap, 427, 319

\bibitem[{Bellot~Rubio {et~al.}(2003)Bellot~Rubio, Balthasar, Collados, \&
  Schlichenmaier}]{2003A&A...403L..47B}
Bellot~Rubio, L.~R., Balthasar, H., Collados, M., \& Schlichenmaier, R. 2003,
  \aap, 403, L47

\bibitem[{Bumba(1960)}]{bumba60}
Bumba, V. 1960, Izv. Krim. Astro. Obs., 23

\bibitem[{Cabrera~Solana {et~al.}(2006)Cabrera~Solana, Bellot~Rubio, Beck, \&
  del Toro~Iniesta}]{2006ApJ...649L..41C}
Cabrera~Solana, D., Bellot~Rubio, L.~R., Beck, C., \& del Toro~Iniesta, J.~C.
  2006, \apjl, 649, L41

\bibitem[{Cabrera~Solana {et~al.}(2007)Cabrera~Solana, Bellot~Rubio, Beck, \&
  del Toro~Iniesta}]{2007A&A...475.1067C}
---. 2007, \aap, 475, 1067

\bibitem[{de~la Cruz~Rodr\'{i}guez {et~al.}(2015)de~la Cruz~Rodr\'{i}guez,
  L\"{o}fdahl, S\"{u}tterlin, Hillberg, \& Rouppe van~der
  Voort}]{2015A&A...573A..40D}
de~la Cruz~Rodr\'{i}guez, J., L\"{o}fdahl, M.~G., S\"{u}tterlin, P., Hillberg,
  T., \& Rouppe van~der Voort, L. 2015, \aap, 573, A40

\bibitem[{{Deinzer}(1965)}]{1965ApJ...141..548D}
{Deinzer}, W. 1965, \apj, 141, 548

\bibitem[{del Toro~Iniesta {et~al.}(2001)del Toro~Iniesta, Bellot~Rubio, \&
  Collados}]{2001ApJ...549L.139D}
del Toro~Iniesta, J.~C., Bellot~Rubio, L.~R., \& Collados, M. 2001, \apjl, 549,
  L139

\bibitem[{Esteban~Pozuelo {et~al.}(2015)Esteban~Pozuelo, Bellot~Rubio, \& de~la
  Cruz~Rodr\'{i}guez}]{2015ApJ...803...93E}
Esteban~Pozuelo, S., Bellot~Rubio, L.~R., \& de~la Cruz~Rodr\'{i}guez, J. 2015,
  \apj, 803, 93

\bibitem[{{Evershed}(1909)}]{1909MNRAS..69..454E}
{Evershed}, J. 1909, \mnras, 69, 454

\bibitem[{Franz \& Schlichenmaier(2009)}]{2009A&A...508.1453F}
Franz, M., \& Schlichenmaier, R. 2009, \aap, 508, 1453

\bibitem[{Franz \& Schlichenmaier(2013)}]{2013A&A...550A..97F}
---. 2013, \aap, 550, A97

\bibitem[{Ichimoto {et~al.}(2007)Ichimoto, Shine, Lites, Kubo, Shimizu,
  Suematsu, Tsuneta, Katsukawa, Tarbell, Title, Nagata, Yokoyama, \&
  Shimojo}]{2007PASJ...59S.593I}
Ichimoto, K., Shine, R.~A., Lites, B., {et~al.} 2007, \pasj, 59, S593

\bibitem[{Kosugi {et~al.}(2007)Kosugi, Matsuzaki, Sakao, Shimizu, Sone,
  Tachikawa, Hashimoto, Minesugi, Ohnishi, Yamada, Tsuneta, Hara, Ichimoto,
  Suematsu, Shimojo, Watanabe, Shimada, Davis, Hill, Owens, Title, Culhane,
  Harra, Doschek, \& Golub}]{2007SoPh..243....3K}
Kosugi, T., Matsuzaki, K., Sakao, T., {et~al.} 2007, \solphys, 243, 3

\bibitem[{Leighton {et~al.}(1962)Leighton, Noyes, \&
  Simon}]{1962ApJ...135..474L}
Leighton, R.~B., Noyes, R.~W., \& Simon, G.~W. 1962, \apj, 135, 474

\bibitem[{{L{\"o}fdahl} \& {Scharmer}(1994)}]{1994A&AS..107..243L}
{L{\"o}fdahl}, M.~G., \& {Scharmer}, G.~B. 1994, \aaps, 107

\bibitem[{Maltby(1964)}]{1964ApNr....8..205M}
Maltby, P. 1964, Astrophysica Norvegica, 8, 205

\bibitem[{Norton {et~al.}(2006)Norton, Graham, Ulrich, Schou, Tomczyk, Liu,
  Lites, L\'{o}pez~Ariste, Bush, Socas-Navarro, \&
  Scherrer}]{2006SoPh..239...69N}
Norton, A.~A., Graham, J.~P., Ulrich, R.~K., {et~al.} 2006, \solphys, 239, 69

\bibitem[{{Rempel}(2012)}]{2012ApJ...750...62R}
{Rempel}, M. 2012, \apj, 750, 62

\bibitem[{Rimmele \& Marino(2006)}]{2006ApJ...646..593R}
Rimmele, T., \& Marino, J. 2006, \apj, 646, 593

\bibitem[{Rimmele(1994)}]{1994A&A...290..972R}
Rimmele, T.~R. 1994, \aap, 290

\bibitem[{Rimmele(1995)}]{1995A&A...298..260R}
---. 1995, \aap, 298, 260

\bibitem[{Ruiz~Cobo \& del Toro~Iniesta(1992)}]{1992ApJ...398..375R}
Ruiz~Cobo, B., \& del Toro~Iniesta, J.~C. 1992, \apj, 398, 375

\bibitem[{Scharmer(2006)}]{2006A&A...447.1111S}
Scharmer, G.~B. 2006, \aap, 447, 1111

\bibitem[{Scharmer {et~al.}(2003a)Scharmer, Bjelksjo, Korhonen, Lindberg, \&
  Petterson}]{2003SPIE.4853..341S}
Scharmer, G.~B., Bjelksjo, K., Korhonen, T.~K., Lindberg, B., \& Petterson, B.
  2003a, in \procspie, Vol. 4853, Innovative Telescopes and Instrumentation for
  Solar Astrophysics, ed. S.~L. {Keil} \& S.~V. {Avakyan}, 341--350

\bibitem[{{Scharmer} {et~al.}(2008){Scharmer}, {Narayan}, {Hillberg}, {de la
  Cruz Rodriguez}, {L{\"o}fdahl}, {Kiselman}, {S{\"u}tterlin}, {van Noort}, \&
  {Lagg}}]{2008ApJ...689L..69S}
{Scharmer}, G.~B., {Narayan}, G., {Hillberg}, T., {et~al.} 2008, \apjl, 689,
  L69

\bibitem[{{Schlichenmaier}(2002)}]{2002AN....323..303S}
{Schlichenmaier}, R. 2002, Astronomische Nachrichten, 323, 303

\bibitem[{{Servajean}(1961)}]{1961AnAp...24....1S}
{Servajean}, R. 1961, Annales d'Astrophysique, 24, 1

\bibitem[{{Severny}(1960)}]{1960IAUS...12..403S}
{Severny}, A.~B. 1960, in IAU Symposium, Vol.~12, Aerodynamic Phenomena in
  Stellar Atmospheres, ed. R.~N. {Thomas}, 403--415

\bibitem[{Shine {et~al.}(1994)Shine, Title, Tarbell, Smith, Frank, \&
  Scharmer}]{1994ApJ...430..413S}
Shine, R.~A., Title, A.~M., Tarbell, T.~D., {et~al.} 1994, \apj, 430, 413

\bibitem[{Sobotka {et~al.}(1999)Sobotka, Brandt, \&
  Simon}]{1999A&A...348..621S}
Sobotka, M., Brandt, P.~N., \& Simon, G.~W. 1999, \aap, 348, 621

\bibitem[{Sobotka \& S\"{u}tterlin(2001)}]{2001A&A...380..714S}
Sobotka, M., \& S\"{u}tterlin, P. 2001, \aap, 380, 714

\bibitem[{Stellmacher \& Wiehr(1971)}]{1971SoPh...18..220S}
Stellmacher, G., \& Wiehr, E. 1971, \solphys, 18, 220

\bibitem[{{Thomas}(1988)}]{1988ApJ...333..407T}
{Thomas}, J.~H. 1988, \apj, 333, 407

\bibitem[{Thonhofer {et~al.}(2015)Thonhofer, Bellot~Rubio, Utz, Hanslmeier, \&
  Jur\c{c}\'{a}k}]{2015IAUS..305..251T}
Thonhofer, S., Bellot~Rubio, L.~R., Utz, D., Hanslmeier, A., \& Jur\c{c}\'{a}k,
  J. 2015, in IAU Symposium, Vol. 305, Polarimetry, ed. K.~N. {Nagendra},
  S.~{Bagnulo}, R.~{Centeno}, \& M.~{Jes{\'u}s Mart{\'{\i}}nez Gonz{\'a}lez},
  251--256

\bibitem[{Title {et~al.}(1989)Title, Tarbell, Topka, Ferguson, Shine, \&
  Team}]{1989ApJ...336..475T}
Title, A.~M., Tarbell, T.~D., Topka, K.~P., {et~al.} 1989, \apj, 336, 475

\bibitem[{van Noort(2012)}]{2012A&A...548A...5V}
van Noort, M. 2012, \aap, 548, A5

\bibitem[{van Noort {et~al.}(2013)van Noort, Lagg, Tiwari, \&
  Solanki}]{2013A&A...557A..24V}
van Noort, M., Lagg, A., Tiwari, S.~K., \& Solanki, S.~K. 2013, \aap, 557, A24

\bibitem[{van Noort {et~al.}(2005)van Noort, Rouppe van~der Voort, \&
  L\"{o}fdahl}]{2005SoPh..228..191V}
van Noort, M., Rouppe van~der Voort, L., \& L\"{o}fdahl, M.~G. 2005, \solphys,
  228, 191

\bibitem[{van Noort \& Rouppe van~der Voort(2008)}]{2008A&A...489..429V}
van Noort, M.~J., \& Rouppe van~der Voort, L.~H.~M. 2008, \aap, 489, 429

\bibitem[{Wang \& Zirin(1992)}]{1992SoPh..140...41W}
Wang, H., \& Zirin, H. 1992, \solphys, 140, 41

\bibitem[{Wiehr(1995)}]{1995A&A...298L..17W}
Wiehr, E. 1995, \aap, 298, L17

\end{thebibliography}

\end{document}